\documentclass[pdflatex,sn-mathphys-num]{sn-jnl}
\usepackage{geometry}
\usepackage{graphicx}%
\usepackage{multirow}%
\usepackage{amsmath,amssymb,amsfonts}%
\usepackage{amsthm}%
\usepackage{mathrsfs}%
\usepackage[title]{appendix}%
\usepackage{xcolor}%
\usepackage{textcomp}%
\usepackage{manyfoot}%
\usepackage{booktabs}%
\usepackage{algorithm}%
\usepackage{algorithmicx}%
\usepackage{algpseudocode}%
\usepackage{listings}%

\usepackage[scr=rsfs]{mathalpha}
\usepackage{longtable}
\usepackage{float} 
\usepackage{lmodern}
\usepackage{array}%

\theoremstyle{thmstyleone}%
\theoremstyle{thmstyletwo}%

\theoremstyle{thmstylethree}%

\begin{document}

\title[Benchmarking Universal Interatomic Potentials on Elemental Systems]{
Benchmarking Universal Machine Learning Interatomic Potentials on Elemental Systems}

%
%

\author[1,2]{\fnm{Hossein} \sur{Tahmasbi}}
\author[1,2]{\fnm{Andreas} \sur{Kn\"upfer}}
\author[1,2]{\fnm{Thomas D.} \sur{K\"uhne}}
\author*[1,2]{\fnm{Hossein} \sur{Mirhosseini}}\email{h.mirhosseini@hzdr.de}
\affil[1]{Center for Advanced Systems Understanding (CASUS), D-02826 G\"orlitz, Germany}

\affil[2]{Helmholtz-Zentrum Dresden-Rossendorf (HZDR), D-01328 Dresden, Germany}


\abstract{The rapid emergence of universal Machine Learning Interatomic Potentials (uMLIPs) has transformed materials modeling. However, a comprehensive understanding of their generalization behavior across configurational space remains an open challenge.
In this work, we introduce a benchmarking framework to evaluate both the equilibrium and far-from-equilibrium performance of state-of-the-art uMLIPs, including three MACE-based models, MatterSim, and PET-MAD. Our assessment utilizes Equation-of-State (EOS) tests to evaluate near-equilibrium properties, such as bulk moduli and equilibrium volumes, alongside extensive Minima Hopping (MH) structural searches to probe the global Potential Energy Surface (PES). Here, we assess universality within the fundamental limit of unary (elemental) systems, which serve as a necessary baseline for broader chemical generalization and provide a framework that can be systematically extended to multicomponent materials. 
We find that while most models exhibit high accuracy in reproducing equilibrium volumes for transition metals, significant performance gaps emerge in alkali and alkaline earth metal groups. Crucially, our MH results reveal a decoupling between search efficiency and structural fidelity, highlighting that smoother learned PESs do not necessarily yield more accurate energetic landscapes.

}


\maketitle

\section{Introduction}
Accurate predictions of interatomic interactions are fundamental to understanding the structure–property relationships that govern materials behavior. 
Conventional \textit{ab initio} approaches, most notably density functional theory (DFT), have long served as the computational workhorse for predicting potential energy surfaces (PES) with reliable accuracy. However, their substantial computational cost severely limits the accessible length and time scales, particularly for large or complex systems~\cite{Hafner2006}. In response to these limitations, machine-learning interatomic potentials (MLIPs) have emerged as a powerful alternative, reproducing near–DFT accuracy at drastically reduced computational expense~\cite{Mueller2020,Unke2021}. By efficiently approximating high-dimensional PESs, MLIPs enable large-scale atomistic simulations and high-throughput materials screening that are otherwise intractable with first-principles methods~\cite{Schmidt2019,Curtarolo2013}. This capability is instrumental in predicting novel crystal structures and significantly accelerating the discovery and optimization of new materials, as indicated by recent applications to increasingly complex chemical compositions and energy landscapes~\cite{Deringer2021,Tahmasbi2021,Rosen2021,Merchant2023,Tahmasbi2024}.

The framework for modern MLIPs was established by several foundational methods that introduced sophisticated descriptors for the local atomic environment. Notable among these are the Behler-Parrinello neural network potentials~\cite{Behler2007}, Gaussian Approximation Potentials (GAPs)~\cite{Bartk2010}, Spectral Neighborhood Analysis Potentials (SNAPs)~\cite{Thompson2015}, Moment Tensor Potentials (MTPs)~\cite{Shapeev2016}, and the Atomic Cluster Expansion (ACE) framework~\cite{Drautz2019}. 

Recent progress has shifted the field from system-specific MLIPs toward universal MLIPs (uMLIPs) foundation models trained to describe broad regions of chemical and structural space with a single potential. By leveraging graph neural networks (GNNs) and equivariant message-passing architectures, these models have demonstrated impressive accuracy and transferability. Key examples of successful uMLIPs include MEGNet~\cite{Chen2019}, Nequip~\cite{Batzner2022}, MACE~\cite{mace2022,mace2024}, Allegro~\cite{Musaelian2023}, M3GNet~\cite{Chen2022}, MatterSim~\cite{Mattersim}, and PET-MAD~\cite{Mazitov2025}. These models consistently demonstrate robustness across diverse chemical environments, paving the way for autonomous, high-throughput materials discovery~\cite{Deng2023,Takamoto2022,Batatia2025,Yang2024,Yang2025E2GNN}.

Nevertheless, ensuring the reliability of uMLIPs across the immense diversity of atomic environments sampled in real-world materials remains an open challenge. Progress in this area therefore depends on two pillars: the availability of high-quality datasets that comprehensively span configurational space and the development of rigorous benchmarks capable of exposing limitations across both locally stable configurations and more complex regions of the energy landscape. While automated workflows, high-throughput sampling, and active learning through implementations such as \texttt{PyFLAME}~\cite{mirhosseini2021}, \texttt{wfl}~\cite{Gelinyt2023}, and \texttt{autoplex}~\cite{Liu2025}, are beginning to address the data bottleneck by producing datasets that span both equilibrium and non-equilibrium configurations, systematic benchmarks are essential to assess whether uMLIPs can generalize across distinct physical regimes.

Previous efforts in atomistic machine learning and materials modelling have established a rich landscape for evaluating uMLIPs. Recent work includes model-specific evaluations, such as the benchmarking of CHGNet~\cite{guns2025}, as well as property-focused assessments for phenomena such as phonons~\cite{Loew2025, Bandi2024, Anam2025}. Community-driven platforms like Matbench~\cite{Dunn2020}, Matbench Discovery~\cite{Riebesell2025}, and MLIP Arena~\cite{MLIPArena2025} provide standardized protocols for cross-model comparisons. Complementing these broad evaluations are specialized benchmarks for high-demand applications, ranging from the analysis of inelastic neutron scattering data~\cite{Han2025} to the assessment of supported Cu nanoparticles through the decoupling of energy accuracy and structural exploration~\cite{Xu2025_glopt}, as well as specific material classes like Li–P–S solid-state electrolytes~\cite{Fragapane2025} and metal-organic frameworks (MOFs)~\cite{krassuniversal,Krass2025}. Furthermore, studies on crystal stability emphasize the need for consistent metrics and reproducible protocols~\cite{Riebesell2025}. Consequently, a rigorous evaluation must test uMLIPs against both general standards and stringent physical scenarios to ensure their reliability in real-world scientific applications.

In this work, we introduce a benchmarking framework that evaluates uMLIPs along two complementary axes. First, equation of state (EOS) tests assess the accuracy of near-equilibrium structural properties, including equilibrium volume, bulk modulus, and PES curvature. Second, global optimization, performed here via the Minima Hopping (MH) algorithm~\cite{Goedecker2004,Amsler2010}, explores a model’s ability to navigate complex energy landscapes, recover known global minimum (GM) structures, avoid unphysical minima, and maintain stability even under large structural distortions. Together, these tasks provide a balanced assessment of both local accuracy and global generalization. Unlike property-centric benchmarks, the present work emphasizes direct exploration of the learned energy landscape, enabling identification of failure modes that are not captured by standard accuracy metrics alone. This dual-axis evaluation enables a detailed understanding of model robustness and highlights pathways for improving data coverage, model architectures, and training protocols of next-generation foundation interatomic potentials.

We apply this framework to a selection of state-of-the-art architectures, specifically MatterSim~\cite{Mattersim}, PET-MAD~\cite{Mazitov2025}, and two MACE variants~\cite{Batatia2025foundation} pre-trained on MATPES~\cite{Kaplan2025} and OMAT~\cite{omat2024}, respectively, as well as a custom MACE model that has been trained using our own generated unary dataset. 

As a foundational step toward evaluating universality, we focus on unary (elemental) systems. While unary systems represent a reduced chemical space compared to multicomponent materials, they span a wide diversity of bonding motifs, including metallic, covalent, and semiconducting interactions, and expose fundamental limitations in PES smoothness, curvature, and energetic ordering. Accurate performance in this regime is therefore a necessary, though not sufficient, condition for broader universality. By establishing accuracy, stability, and physical consistency behavior across a diverse set of elemental environments, we provide a rigorous baseline for assessing universality and identifying pathways for further improvement. 

\section{Benchmark tasks}
\subsection{EOS}\label{subs:bt_eos}
EOS calculations quantify how total energy varies with changes in volume around the equilibrium geometry. These curves encode key mechanical and structural properties such as equilibrium volume ($V_0$), bulk modulus ($B_0$), and compressibility. Accurate reproduction of EOS behavior requires a model to capture fine variations in bonding interactions and the correct curvature of the PES near the ground state.

Thus, EOS benchmarks directly evaluate the accuracy of predicted equilibrium lattice parameters as well as the curvature of the PES around equilibrium. Since equilibrium properties are foundational for predicting phase stability, mechanical response, and thermodynamic behavior, EOS benchmarking is indispensable for testing uMLIPs intended for general-purpose materials modeling.

For the EOS benchmarks, we compare the MLIP predictions against reference DFT data computed using the exchange–correlation functional consistent with the training of each model. Specifically, the EOS results for PET-MAD~\cite{Mazitov2025} are evaluated against DFT calculations performed with PBEsol~\cite{Perdew2008}, whereas the remaining uMLIPs are benchmarked against PBE~\cite{Perdew1996} reference data. In the case of the MACE-OMAT model~\cite{Batatia2025foundation}, Hubbard-U corrections were applied for the transition metals Co, Cr, Fe, Mn, Mo, Ni, and V. Because these elements were computed with PBE+U rather than plain PBE, their EOS results are not directly comparable to the PBE reference dataset.

\subsubsection{EOS Metrics}
\textbf{Equilibrium volume}:
EOS benchmarks quantify the near-equilibrium structural accuracy of a model. A key measure is the equilibrium volume error
\begin{equation}
    \Delta V_0  (\%)= 200 * \frac{V_0^{MLIP} - V_0^{DFT}} {V_0^{MLIP} + V_0^{DFT}},
\end{equation}
which evaluates the agreement between MLIP-predicted ($V^{MLIP}$) and DFT-calculated equilibrium volumes ($V^{DFT}$). 

\textbf{Bulk modulus}:
The shape of the energy–volume curve provides both qualitative and quantitative insight into whether the model captures the correct curvature of the potential energy surface around the ground state. When assessed, properties such as the bulk modulus further probe higher-order derivatives of the PES, reflecting the material’s mechanical stiffness. The bulk modulus error defined as 
\begin{equation}
    \Delta B_0  (\%)= 200 * \frac{B_0^{MLIP} - B_0^{DFT}} {B_0^{MLIP} + B_0^{DFT}},
\end{equation}
evaluates the agreement between MLIP-predicted bulk modulus ($B_0^{MLIP}$) and DFT-calculated bulk modulus ($B_0^{DFT}$). 

Element-wise performance maps reveal systematic trends or deficiencies across the periodic table, thereby enabling a broad assessment of transferability. Collectively, these metrics provide a high-resolution view of the fidelity of uMLIP predictions in the equilibrium regime.

\subsection{Global Optimization}
In contrast to EOSs, global optimization provides a fundamentally different benchmark by assessing how well an MLIP guides global structure search. We employed MH, which explores the PES by iteratively performing high-temperature molecular dynamics moves and local relaxations, sampling a broad range of metastable and high-energy configurations~\cite{Goedecker2004}.

A reliable uMLIP must therefore assign meaningful forces and energies across diverse regions of configuration space—not just near equilibrium. MH challenges models in several critical ways. First, a reliable uMLIP must maintain a smooth and physically consistent PES even under large structural distortions. It must also correctly rank local minima and successfully recover known ground-state phases, ensuring that the energetic landscape is represented with sufficient fidelity. In addition, the model must avoid unphysical minima or numerical artifacts that can emerge during extensive PES exploration. Finally, robust MH performance requires the model to support efficient structural relaxation, exhibiting low rates of failed or non-convergent optimization steps.

\subsubsection{MH Metrics} \label{subsec:mhm}
The MH benchmarks evaluate the quality of PES exploration and the reliability of global optimization using a given MLIP. The energetic ordering relative to DFT for the discovered local minima assesses whether the model preserves the correct ordering of low-energy structures. The ability to recover known GM and metastable local minima phases further tests whether essential features of the PES are reproduced.  Moreover, the diversity and symmetry of the discovered minima reflect how effectively the potential explores chemically relevant portions of configuration space.

To quantify these aspects in a consistent manner, we introduce two central metrics: the recovery score ($R$) and pairwise ordering accuracy (POA). The recovery score measures whether a MLIP identifies the correct low-energy configurations, while the POA measures how well it assigns relative energies to the configurations it discovers. Together, they capture the most critical requirements for generalization beyond equilibrium: correct structural identification and correct energetic hierarchy.

\textbf{Recovery score:}
The recovery metric quantifies how many reference minima from the materials project database (MPDB)~\cite{Jain2013} are identified during MH. The recovery score reflects fidelity to established phases rather than an exhaustive assessment of the full PES and does not penalize the discovery of novel but physically plausible metastable structures. To reflect its central role in phase stability and structure prediction, the ground-state structure is assigned a higher relative weight in the recovery metric.
Specifically, the GM is weighted by 2, whereas all other reference minima receive weight 1. The recovery score is defined as
\begin{equation}
R = \frac{2N_{GM} + N_{other}}{2 + (N_{ref}-1)},
\end{equation}
where $N_{ref}$ is the number of reference minima considered (within 100 meV/atom of the DFT convex hull), $N_{GM}$ indicates whether GM is found (1 if recovered, 0 otherwise), and $N_{other}$ is the number of additional reference minima recovered. The recovery score lies between 0 and 1 and increases with both coverage and correct identification of GM.

\textbf{Pairwise ordering accuracy:}
To quantify how reliably a MLIP reproduces the relative stability of structures, we employ the pairwise ordering accuracy (POA). This metric evaluates whether the model assigns higher or lower energies to structural pairs in the same way as DFT. For each unordered pair of matched minima $(i,j)$ with DFT energies $E^{\mathrm{DFT}}_i$ and $E^{\mathrm{DFT}}_j$ and MLIP energies $E^{\mathrm{MLIP}}_i$ and $E^{\mathrm{MLIP}}_j$, we define an indicator
\begin{equation}
\delta_{ij} =
\begin{cases}
1, & \text{if } \operatorname{sign}\!\left(E^{\mathrm{MLIP}}_i - E^{\mathrm{MLIP}}_j\right)
     = \operatorname{sign}\!\left(E^{\mathrm{DFT}}_i - E^{\mathrm{DFT}}_j\right), \\[4pt]
0, & \text{otherwise}.
\end{cases}
\end{equation}

The pairwise ordering accuracy is then computed as
\begin{equation}
\mathrm{POA} = 
\frac{\sum_{i<j} \delta_{ij}}{\binom{n}{2}},
\end{equation}
where $n$ is the number of matched minima. POA ranges from $0$ to $1$, with higher values indicating a larger fraction of correctly ordered pairs and therefore a more faithful reconstruction of the underlying energy hierarchy. Because the metric requires at least one comparable pair, we report POA only when $n \ge 2$; when fewer matches occur, we assign the uninformative fallback value $\mathrm{POA} = 0.0$. 
Because many elemental polymorphs differ by only a few meV/atom, POA represents an intentionally stringent metric that often probes energy differences below the intrinsic resolution of current uMLIPs. Low POA values should therefore be interpreted as reflecting the challenge of fine energetic ordering rather than complete PES failure.

\textbf{Robustness metric:}
The distribution of optimization-step counts reflects both the efficiency of structural relaxations and the smoothness of the underlying PES. We introduce an instability metric $I$ defined as the sum of (i) the rate of failed relaxations, those requiring more than 1000 optimization steps, and (ii) the fraction of structures that experience severe and unphysical changes in lattice parameters, including both excessive expansion and collapse during MH. A smaller value of $I$ therefore indicates more reliable and stable relaxation behavior.

\section{Methods}
\subsection{Unary Dataset}
Training data were generated using the high-throughput AiiDA workflow engine~\cite{aiida_1, aiida_2} and our automated dataset-generation plugin, aiida-datagen~\cite{aiida-datagen, mirhosseini2021}. Calculations were performed using CP2K~\cite{Khne2020} interfaced with SIRIUS~\cite{Kozhevnikov2019}, enabling efficient plane-wave ground-state computations. All DFT reference energies and forces were obtained using the PBE exchange–correlation functional~\cite{Perdew1996}.

The dataset consists exclusively of unary systems. We sampled both periodic bulk configurations and finite clusters containing 30–60 atoms. For each element, $\approx$ 400 configurations were generated by combining optimized random bulk geometries with compressed structures down to 75\% of their equilibrium volume, enabling broad sampling of near- and far-from-equilibrium environments. The final dataset includes 26524 atomic configurations, divided into training 23998 (90\%), validation 1263 (5\%), and test 1263 (5\%) sets. All DFT calculations were performed as single-point evaluations on pre-optimized or randomly perturbed structures.

\subsection{uMLIPs}
We trained an uMLIP using the MACE GNN framework~\cite{mace2024, Batatia2025, Batatia2025foundation}. We configured the model architecture with a cutoff radius of 7.0 \AA~ to define local environments, 128 channels for message passing, and messages constructed with maximum spherical symmetry (L) of 1 (max-L = 1). During training, the energy loss function was supplemented with isolated atom energies ($E_0$) to ensure correct dissociation limits. Optimization proceeded with a batch size of 30 (for both training and validation) over 250 epochs. 

The total loss function was a weighted sum of the energy and force mean squared errors. Initially, the energy weight was one and the force weight was $100$. After 75\% of the total epochs (epoch 186), the weights switched to 300 for energy and 100 for forces. This weighting schedule progressively shifts the emphasis from forces to a balanced consideration of energy and forces using the SWA protocol, which dynamically adjusts the weights based on the running standard deviation of forces and energies to maintain balanced contributions. Our final model, MACE-Unary, achieves a mean absolute error (MAE) of $33.3$ meV/atom for energy and a force MAE of $106.7$ meV/\AA~ on the test set. 

In addition to the MACE-Unary-PBE-0 model trained on our generated unary dataset (hereafter referred to as MACE-Unary), we benchmarked several state-of-the-art interatomic potentials. These include MACE-MATPES-PBE-0 trained on the MATPES-PBE dataset (hereafter MACE-MATPES)~\cite{Batatia2025foundation,Kaplan2025}, MACE-OMAT-0 trained on OMAT24 (hereafter MACE-OMAT)~\cite{Batatia2025foundation,omat2024}, MatterSim-v1.0.0-5M (hereafter MatterSim)~\cite{Mattersim}, and PET-MAD version 1.0.2 (hereafter PET-MAD)~\cite{Mazitov2025}.

\subsection{MH}
We employed a Python-based implementation of the MH algorithm~\cite{Krummenacher2024} for efficient global optimization and structure prediction, to explore systematically the PESs of the bulk phases of selected elements. The MH runs were conducted using the ASE python toolkit~\cite{HjorthLarsen2017}, with the previously defined uMLIPs serving as the calculator for energies and forces. 
We explore the energy landscapes of 15 elements across various periods and groups of the periodic table (see Fig.~\ref{fig:heatmap} for the full list). Structural searches were performed using a symmetry tolerance of $0.005$ and 200 minima hopping steps per run. We investigate a range of system sizes, between 8 and 32 atoms, performing 20 independent MH runs for each configuration. A specialized set of system sizes (12, 24, 28, and 50 atoms) was explored for Boron. Each run was initialized from random structures generated with diverse space groups.

After completion of all MH runs, the resulting low-energy structures were compiled for each system and universal model. To ensure robust symmetry identification, space groups were recalculated using a slightly looser symmetry tolerance of 0.01, accounting for the fact that structures relaxed with MLIPs may not exhibit perfectly sharp force and energy gradients required for high-precision symmetry determination.

The predicted structures were then benchmarked against known stable and metastable phases from the MPDB~\cite{Jain2013}. Structural matching was performed using the structure-matching routines implemented in the Pymatgen library~\cite{Ong2013}, with lattice, site, and angular tolerances set to 0.05, 0.05, and 0.5, respectively to identify successful matches between the  identified local minima and the reference structures in the MPDB.

\subsection{EOS}
For each element, a set of prototype crystal structures was selected to sample representative bonding environments and coordination motifs, including simple cubic (SC), body-centered cubic (BCC), face-centered cubic (FCC), and diamond (DIA) structures. These structures are publicly available from our previous work on the verification of DFT implementations~\cite{Bosoni2024}. For each prototype, a sequence of uniformly scaled volumes was generated in the vicinity of the reference equilibrium volume to construct the corresponding energy–volume relationship. EOS predictions from models trained on PBE-based datasets were compared against VASP–PBE reference calculations, while PET-MAD results were benchmarked against SIRIUS–PBEsol reference data, in accordance with the training protocol of each model. 

The energy–volume data were fitted using a third-order Birch–Murnaghan equation of state~\cite{Birch1947,murnaghan1944compressibility}, from which the equilibrium volumes $V_0$  were extracted and compared against the corresponding DFT reference values. EOS curves exhibiting unphysical behavior, such as negative curvature near the minimum, the absence of a well-defined minimum, or numerical instabilities during fitting, were classified as failed and excluded from quantitative analysis.

\section{Results and Discussion}
\subsection{EOS}
EOS calculations provide a rigorous assessment of the ability of a MLIP to reproduce equilibrium structural properties. By mapping the relationship between total energy and volume, EOS benchmarks probe the accuracy of predicted equilibrium lattice parameters and bulk moduli. These quantities are sensitive to the curvature of the potential energy surface in proximity to the ground state. EOS calculations therefore serve as a critical measure of equilibrium accuracy.

\begin{figure}[H]
\centering
\includegraphics[width=0.95\linewidth]{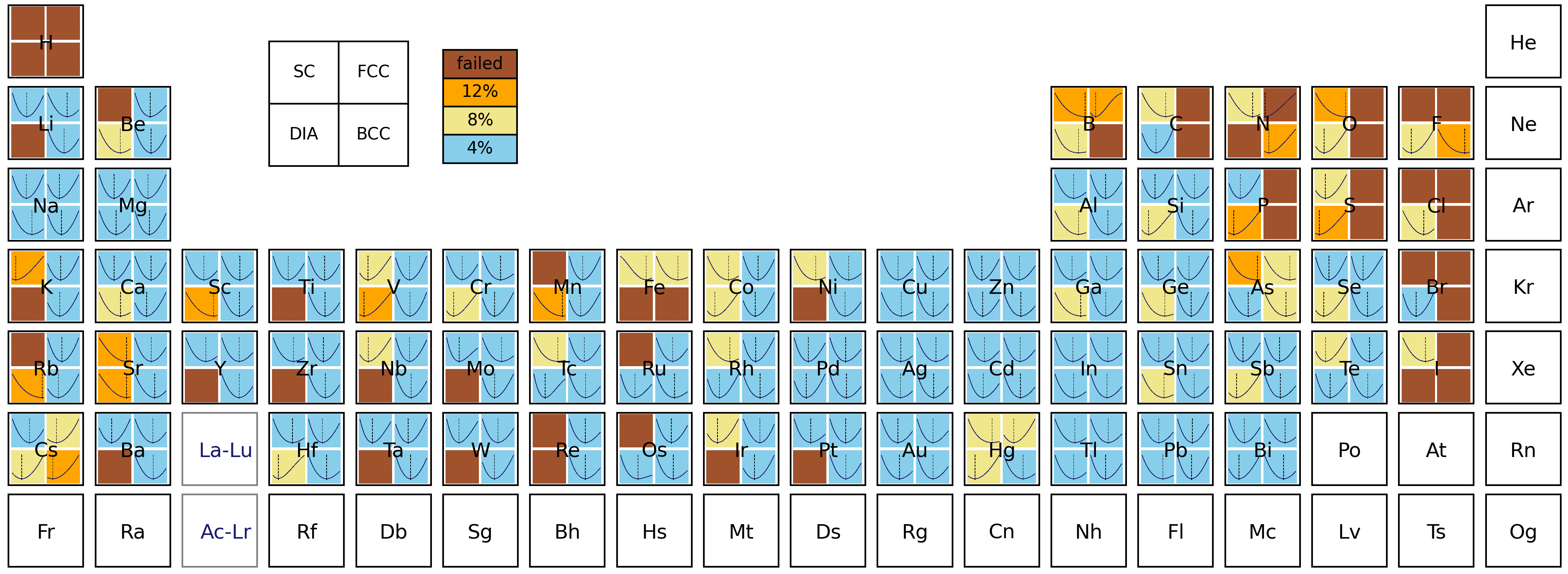}
\caption{
Periodic table map showing the element-wise equilibrium-volume error ($\Delta V_0$) for the MACE-Unary model. Colors indicate the percent deviation relative to DFT reference equilibrium volumes.}
\label{fig:p_t_um}
\end{figure}

\begin{figure}[H]
\centering
\includegraphics[width=0.95\linewidth]{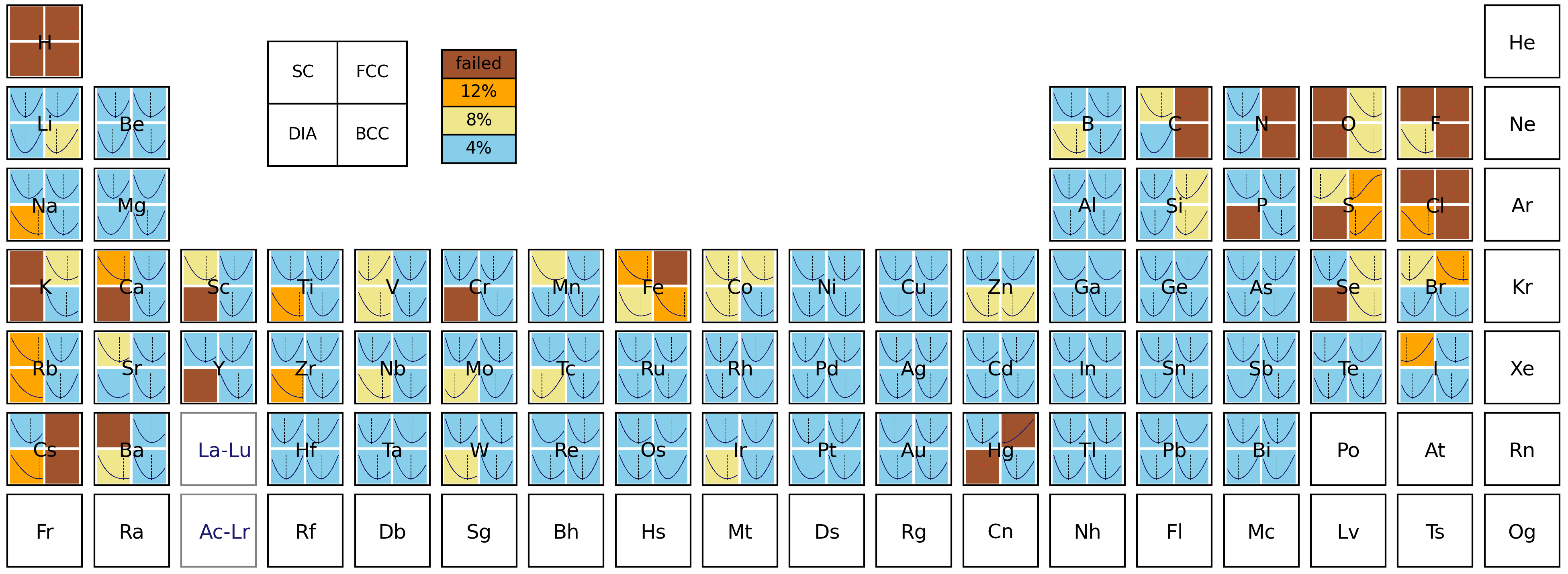}
\caption{
Periodic table map showing the element-wise equilibrium-volume error ($\Delta V_0$) for the MACE-MATPES model. Colors indicate the percent deviation relative to DFT reference equilibrium volumes.}
\label{fig:p_t_matpes}
\end{figure}

\begin{figure}[H]
\centering
\includegraphics[width=0.95\linewidth]{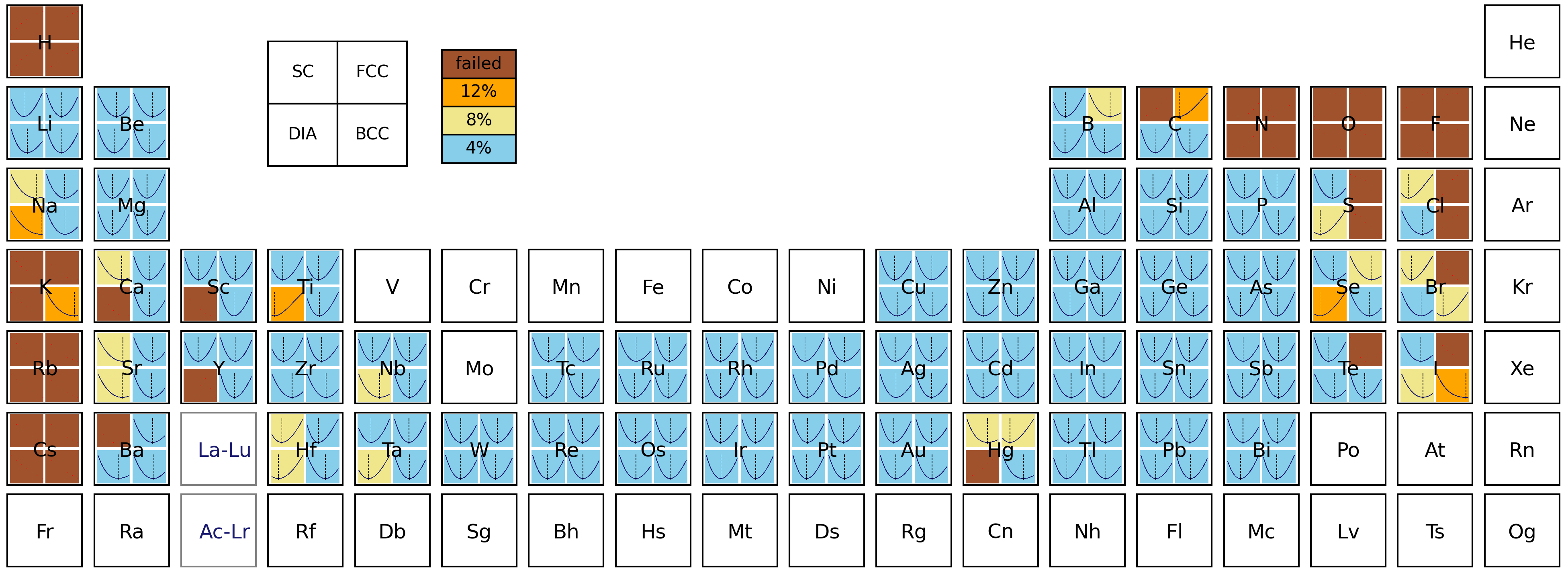}
\caption{
Periodic table map showing the element-wise equilibrium-volume error ($\Delta V_0$) for the MACE-OMAT model. Colors indicate the percent deviation relative to DFT reference equilibrium volumes.}
\label{fig:p_t_omat}
\end{figure}

\begin{figure}[H]
\centering
\includegraphics[width=0.95\linewidth]{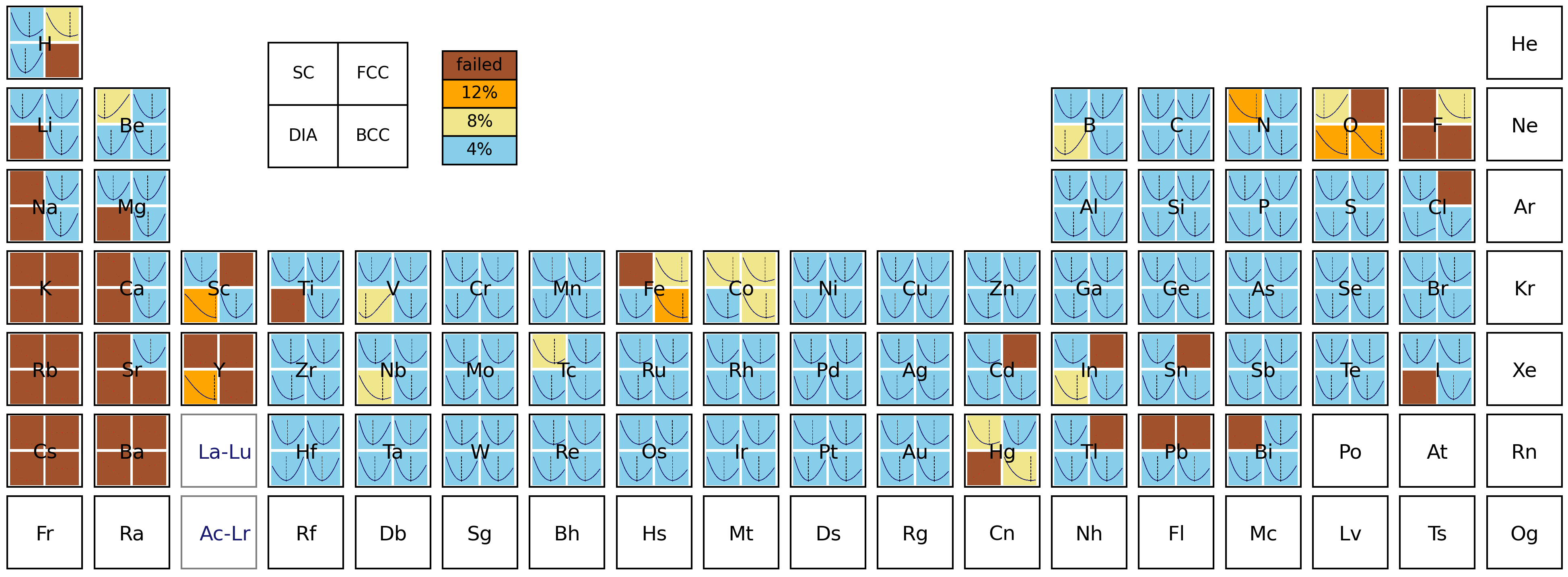}
\caption{
Periodic table map showing the element-wise equilibrium-volume error ($\Delta V_0$) for the MatterSim model. Colors indicate the percent deviation relative to DFT reference equilibrium volumes.}
\label{fig:p_t_ms}
\end{figure}

\begin{figure}[H]
\centering
\includegraphics[width=0.95\linewidth]{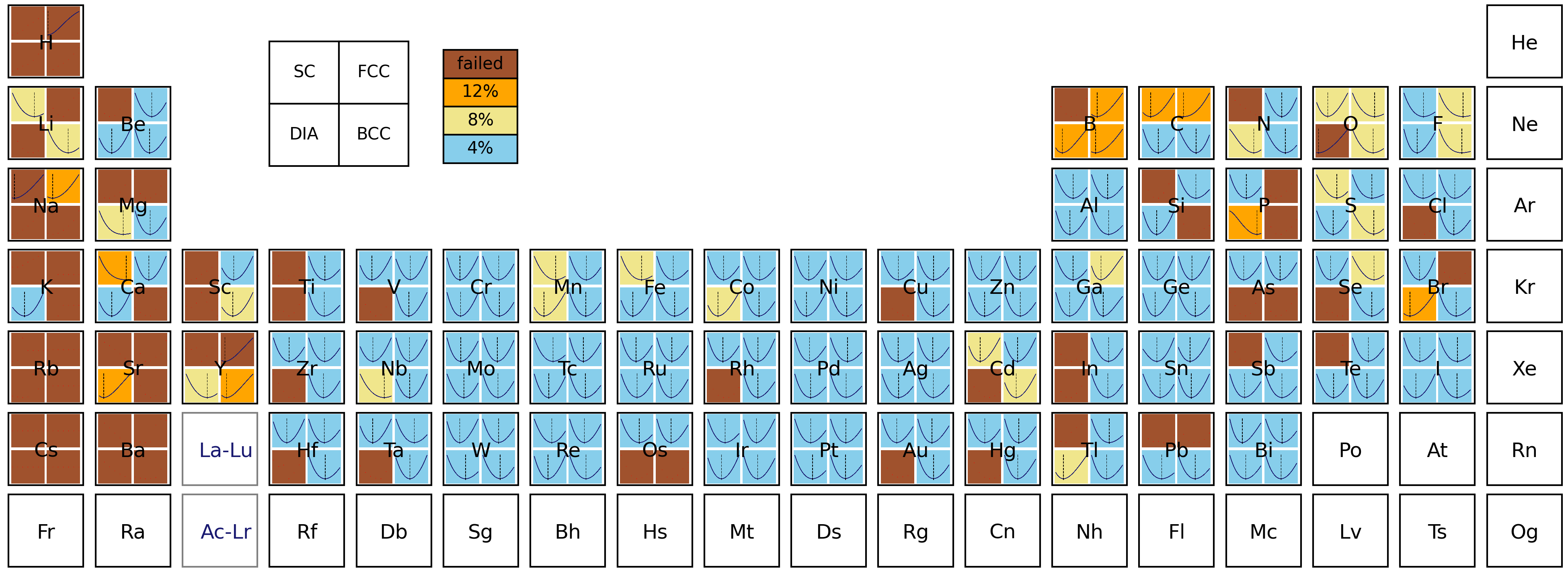}
\caption{
Periodic table map showing the element-wise equilibrium-volume error ($\Delta V_0$) for the PET-MAD model. Colors indicate the percent deviation relative to DFT reference equilibrium volumes.}
\label{fig:p_pm}
\end{figure}

Figs.~\ref{fig:p_t_um}--\ref{fig:p_pm} illustrate the EOS results across the periodic table for five universal models, evaluated for Simple Cubic (SC), Face-Centered Cubic (FCC), Diamond (DIA), and Body-Centered Cubic (BCC) phases.
The colors indicate the percentage of the element-wise equilibrium-volume error ($\Delta V_0$) relative to DFT references. The thresholds are set at 4\% (blue), 10\% (yellow), and 12\% (orange), with brown representing deviations exceeding 12\% or cases where a stable equilibrium volume could not be determined due to unphysical PES curvature.

The models exhibit significant performance variation across different chemical groups. Notably, for transition metal systems, MACE-OMAT and MatterSim consistently yield the most accurate $\Delta V_0$ values. This performance establishes them as the leading models for capturing the physics of these chemically complex elements. Surprisingly, however, MACE-OMAT, MatterSim, and PET-MAD underperform on Group 1 alkali metals when compared to the MACE-Unary and MACE-MATPES models. Furthermore, MatterSim and PET-MAD show the largest deviations within the elements in Group 2 alkaline earth metals.

\begin{figure}[H]
\includegraphics[width=0.19\linewidth]{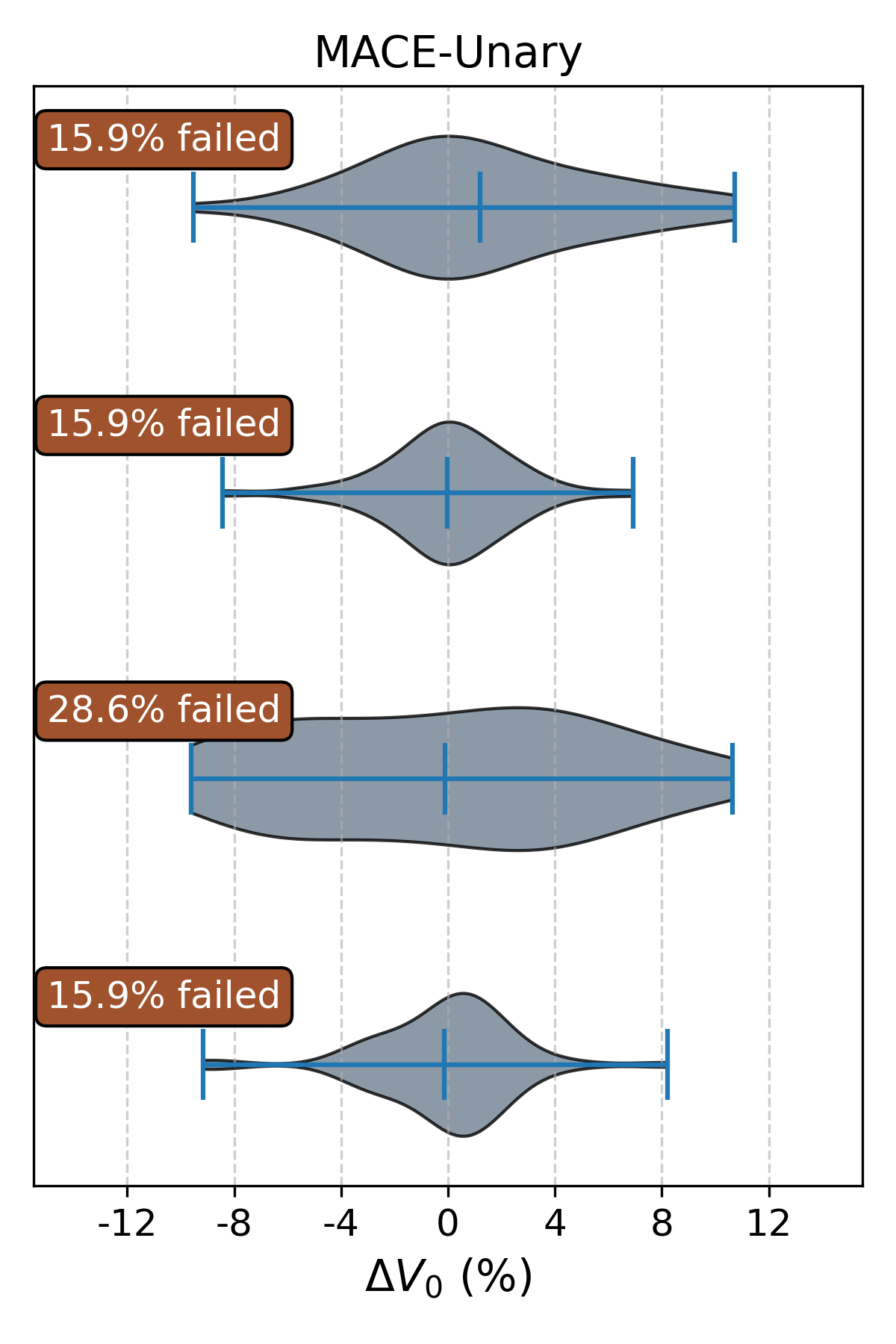}
\includegraphics[width=0.19\linewidth]{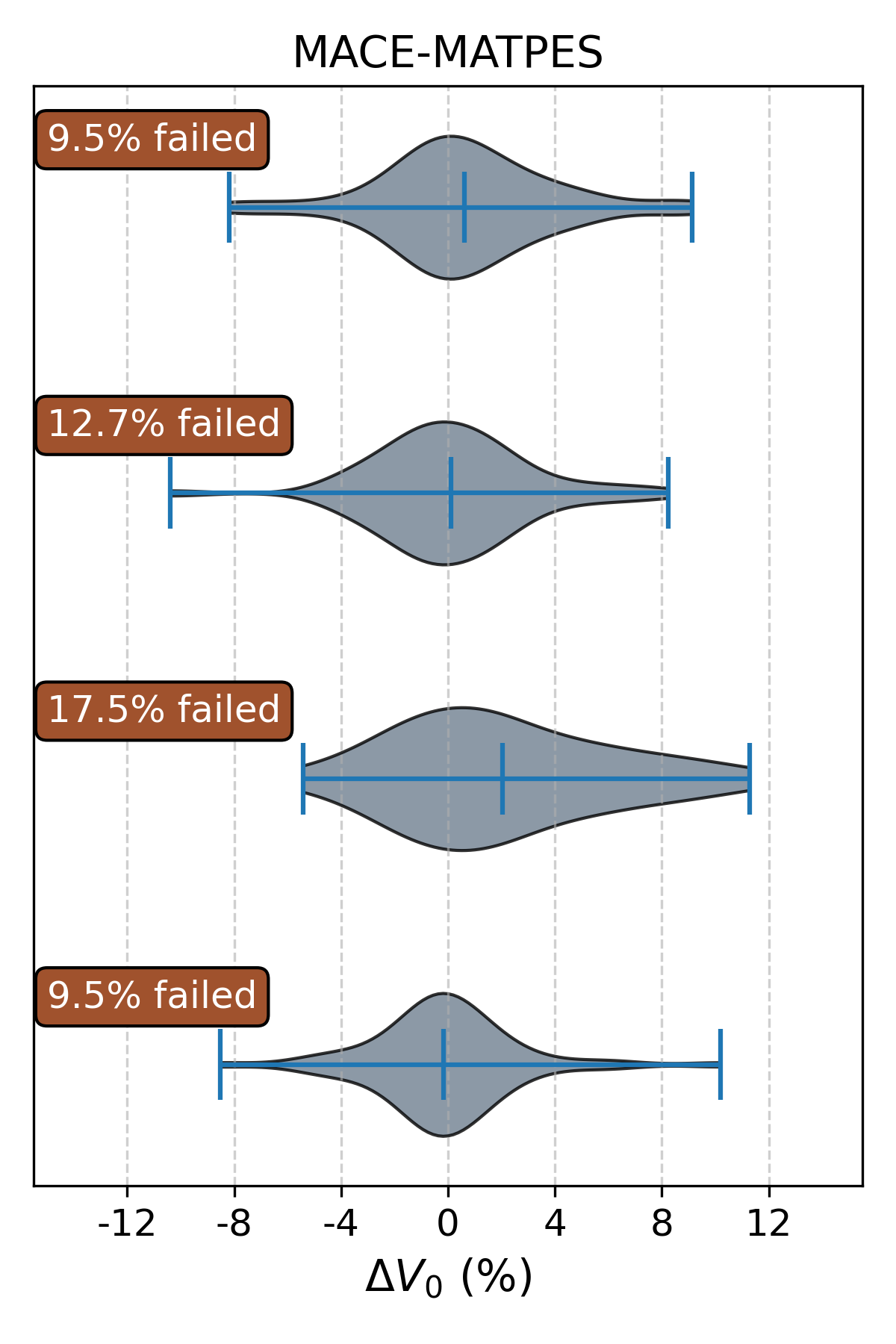}
\includegraphics[width=0.19\linewidth]{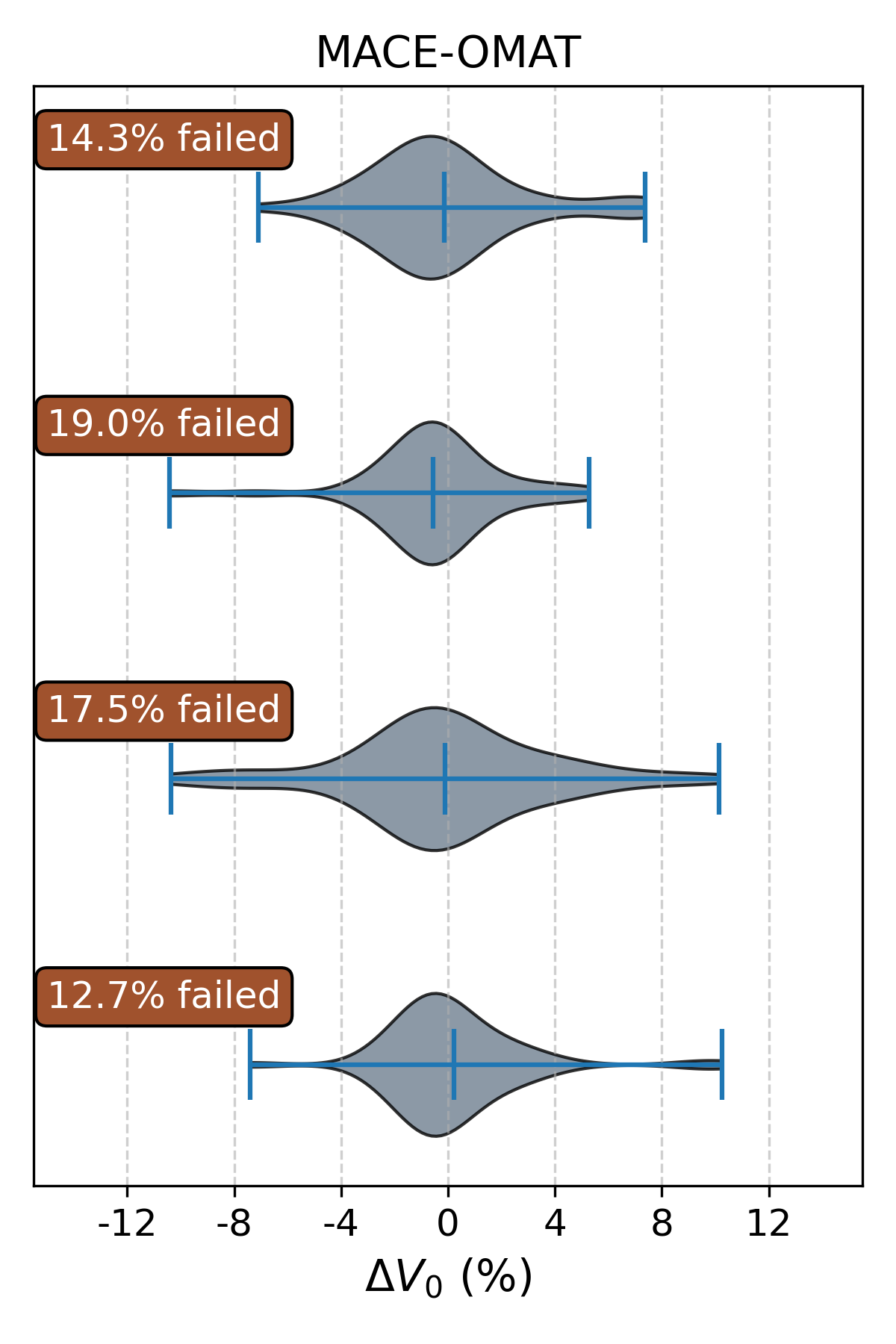}
\includegraphics[width=0.19\linewidth]{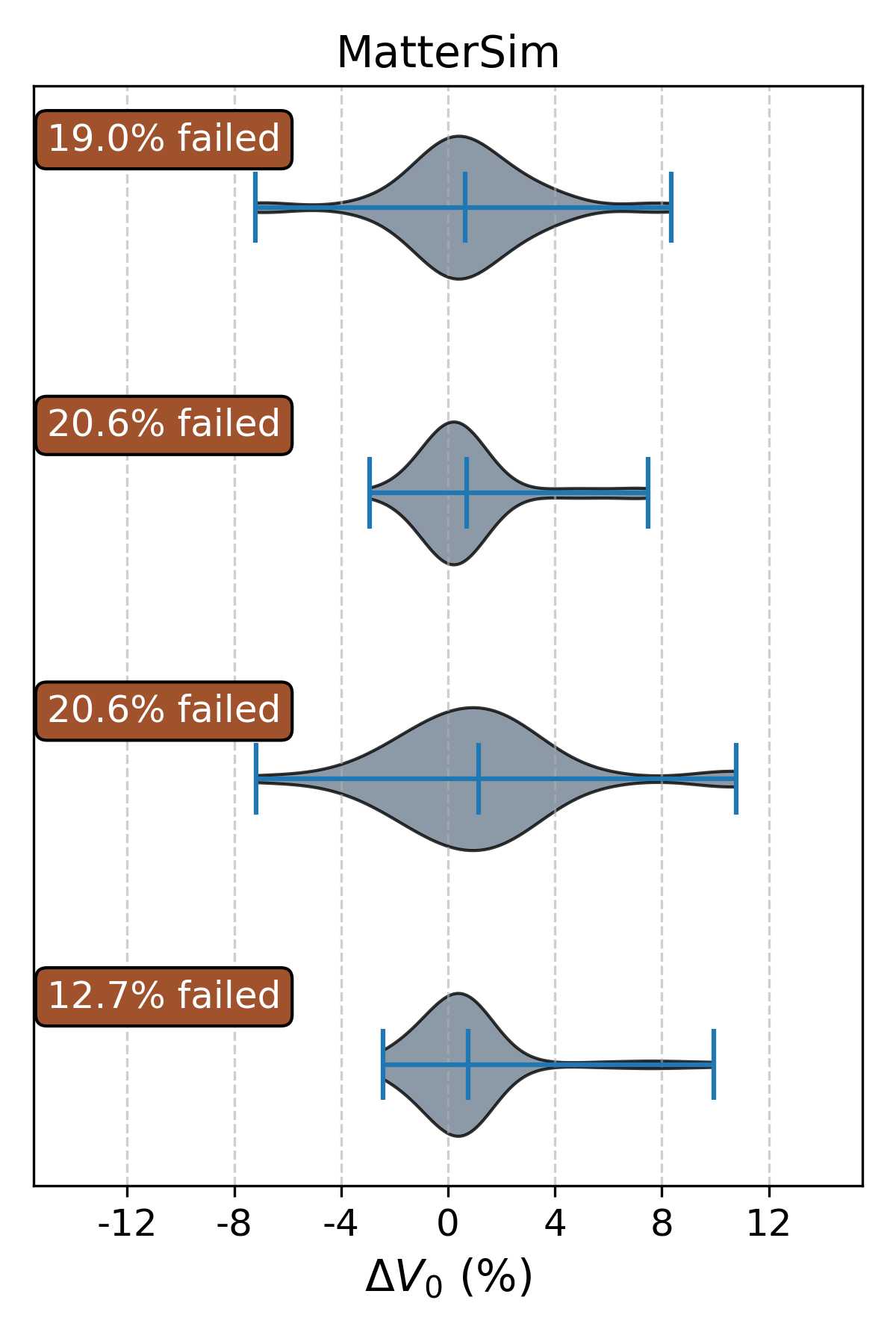}
\includegraphics[width=0.19\linewidth]{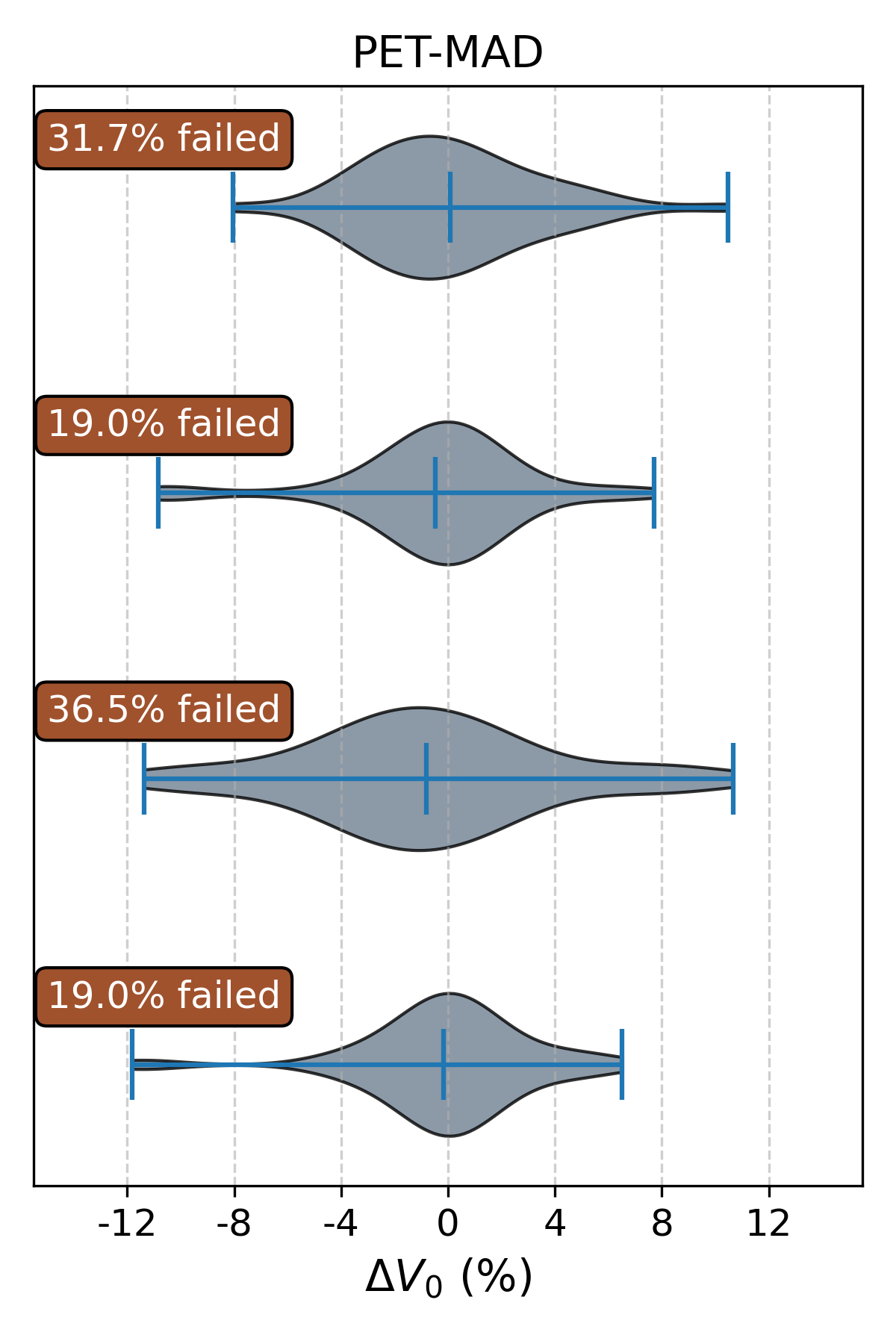}

\includegraphics[width=0.19\linewidth]{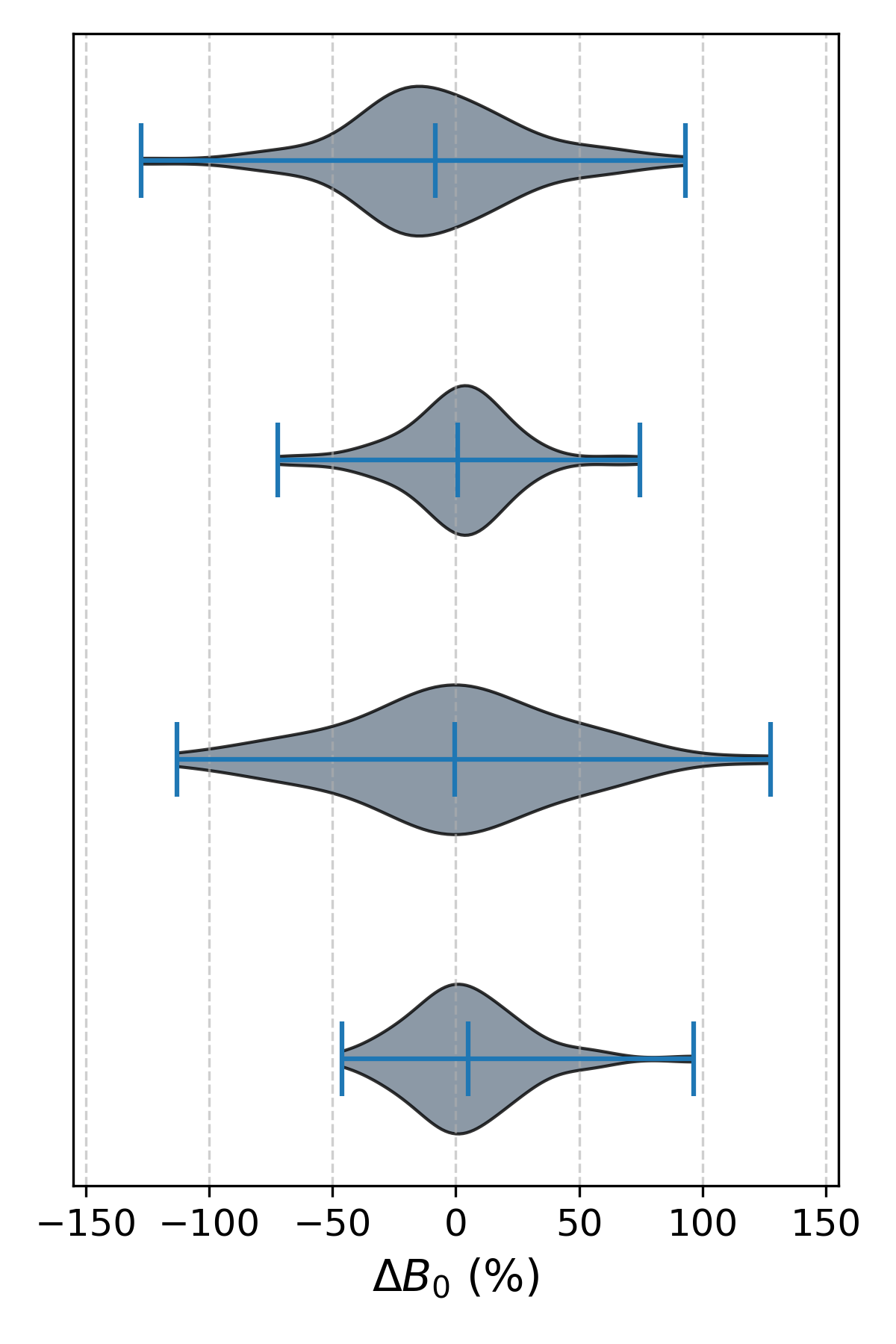}
\includegraphics[width=0.19\linewidth]{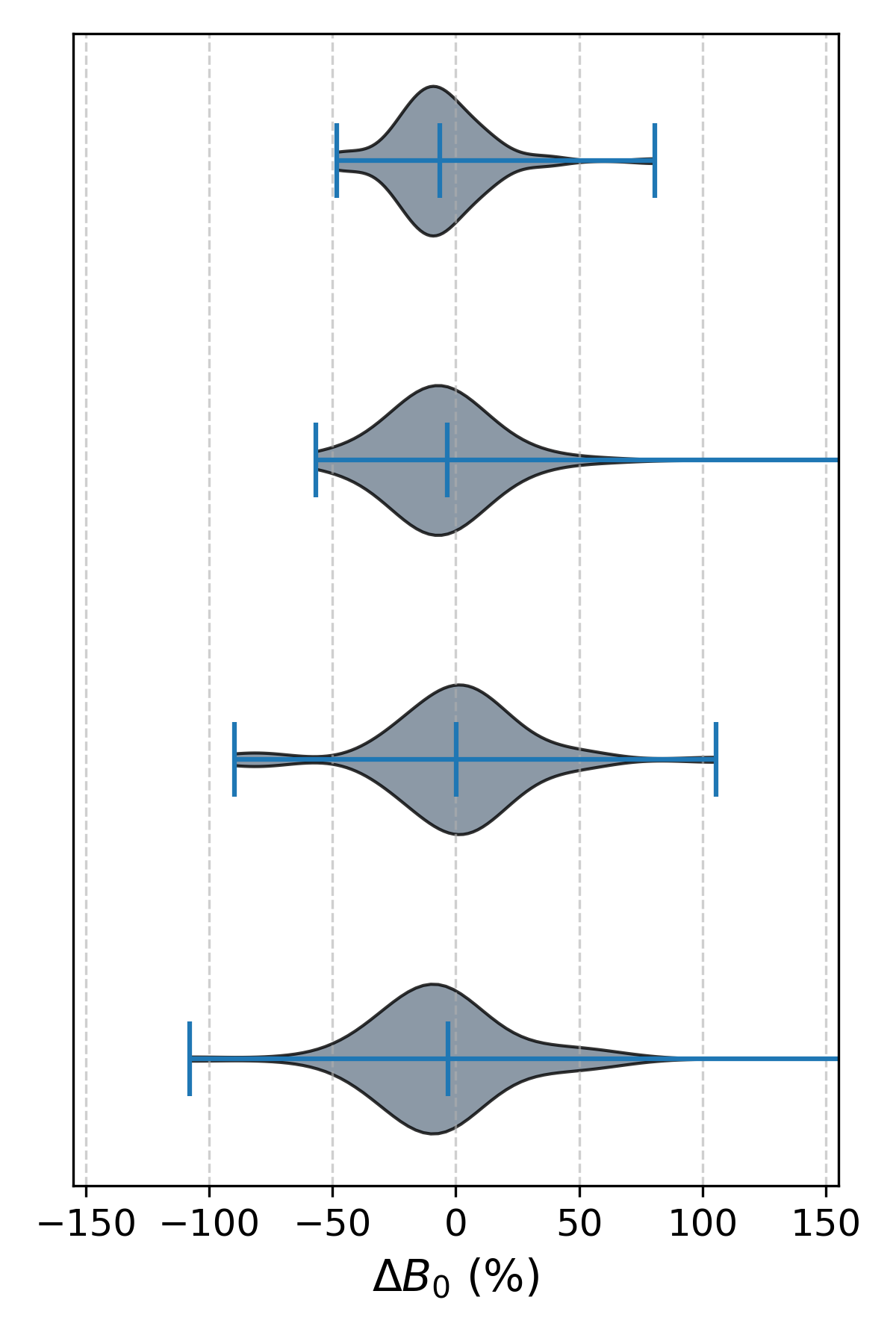}
\includegraphics[width=0.19\linewidth]{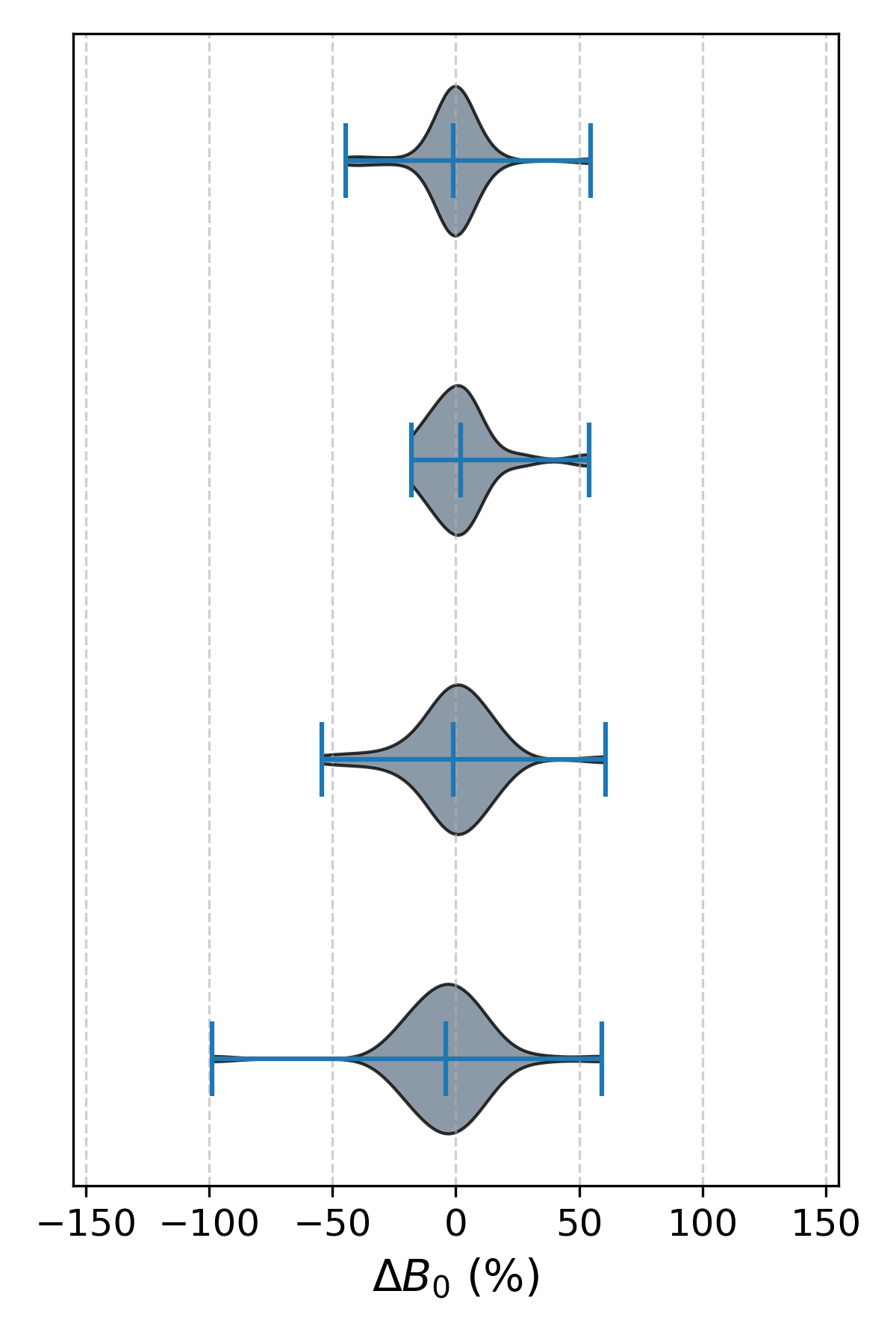}
\includegraphics[width=0.19\linewidth]{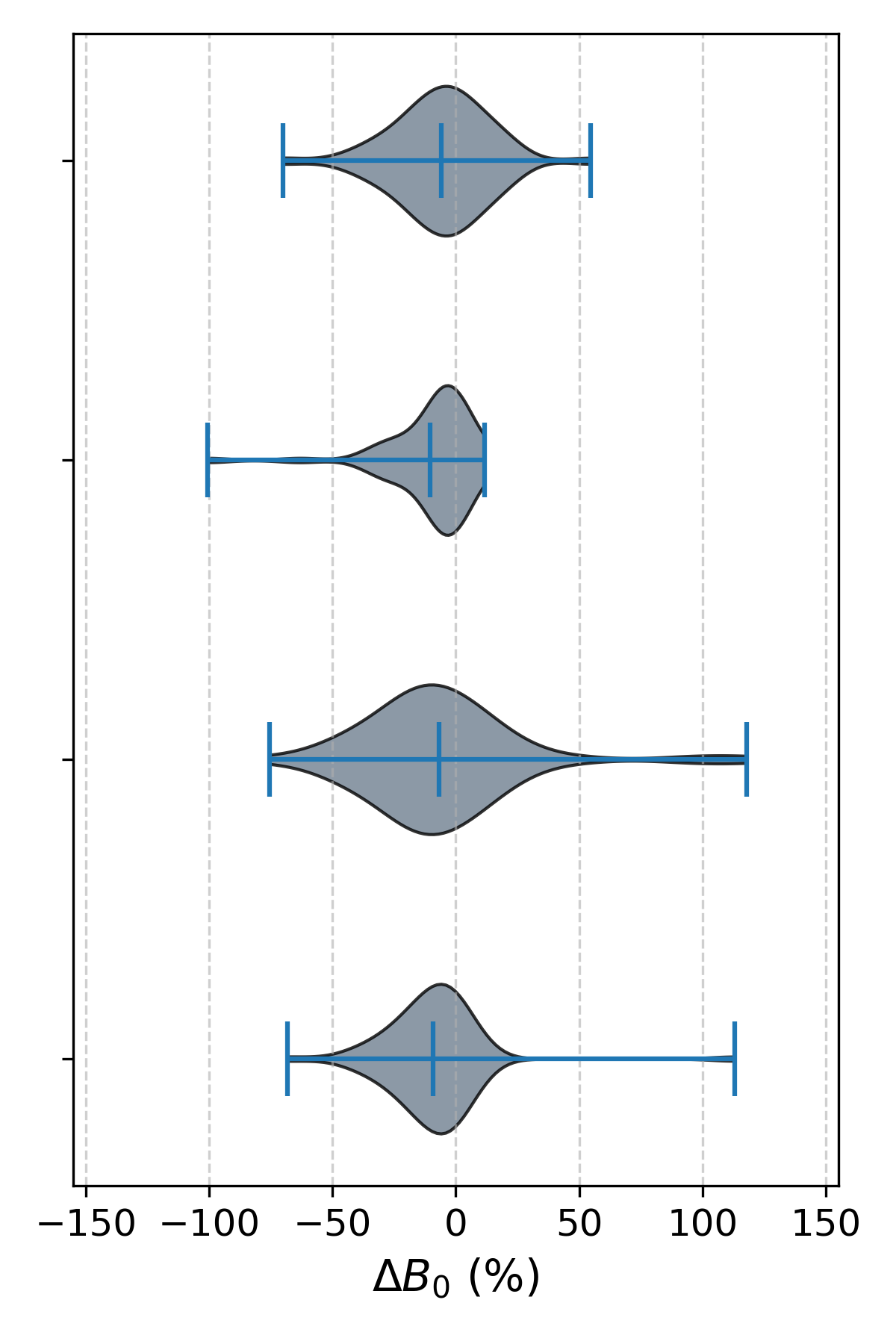}
\includegraphics[width=0.19\linewidth]{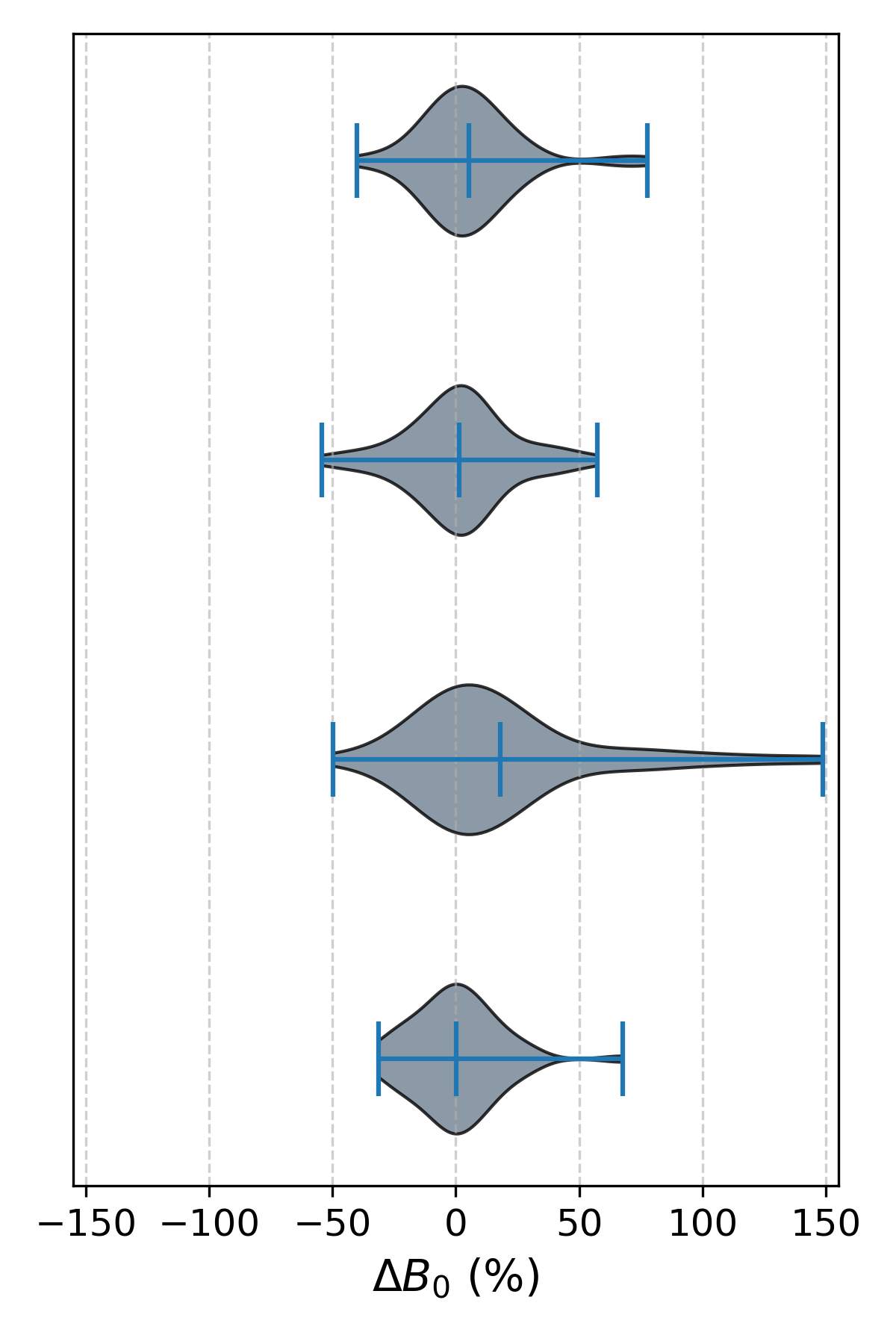}
\caption{
Top: Violin plots depicting  distribution of equilibrium-volume errors ($\Delta V_0$) for all models across four prototype crystal structures (SC, FCC, Diamond, and BCC, ordered from top to bottom). These plots indicate the spread and systematic bias relative to DFT.
Bottom: Violin plots showing the distribution of bulk-modulus errors ($\Delta B_0$) for the same SC, FCC, Diamond, and BCC structures (top to bottom). The plots highlight each model’s ability to reproduce the curvature of the DFT energy–volume curve.}
\label{fig:violin_plots}
\end{figure}

The EOS performance metrics, summarized via the violin plots in Fig.~\ref{fig:violin_plots}, reveal distinct trends in how each model captures the fundamental energy-volume relationship and the PES curvature. Regarding the equilibrium-volume error ($\Delta V_0$), overall both MACE-Unary and PET-MAD exhibit relatively larger deviations, indicating lower precision in reproducing equilibrium lattice parameters. In contrast, MACE-MATPES, MACE-OMAT, and MatterSim demonstrate significantly higher accuracy, showing high fidelity in volume predictions across the SC, FCC, and BCC structures. 

Further differentiation between the uMLIPs emerges in the bulk modulus error ($\Delta B_0$). Here, MACE-Unary shows the weakest performance, struggling to match the DFT-calculated stiffness and curvature. Conversely, MACE-OMAT exhibits the best overall performance, providing the most faithful reproduction of the energy-volume curvature among the tested universal potentials. These findings suggest that while most models can approximate equilibrium points, the ability to maintain PES fidelity under compression and expansion varies significantly, with MACE-OMAT and MatterSim offering the most robust descriptions for structural and mechanical properties.

\subsection{MH}
Benchmarking against MH evaluates whether a potential can assign consistent relative energies and forces to configurations that differ significantly from the equilibrium geometry. This task is especially challenging, as it requires extrapolation across structural motifs and bonding environments not encountered during training. Consequently, MH provides a rigorous test of the generalization capabilities of a potential, as well as its reliability for guiding structure prediction and materials discovery.

\begin{figure}[H]
\centering
\includegraphics[width=\linewidth]{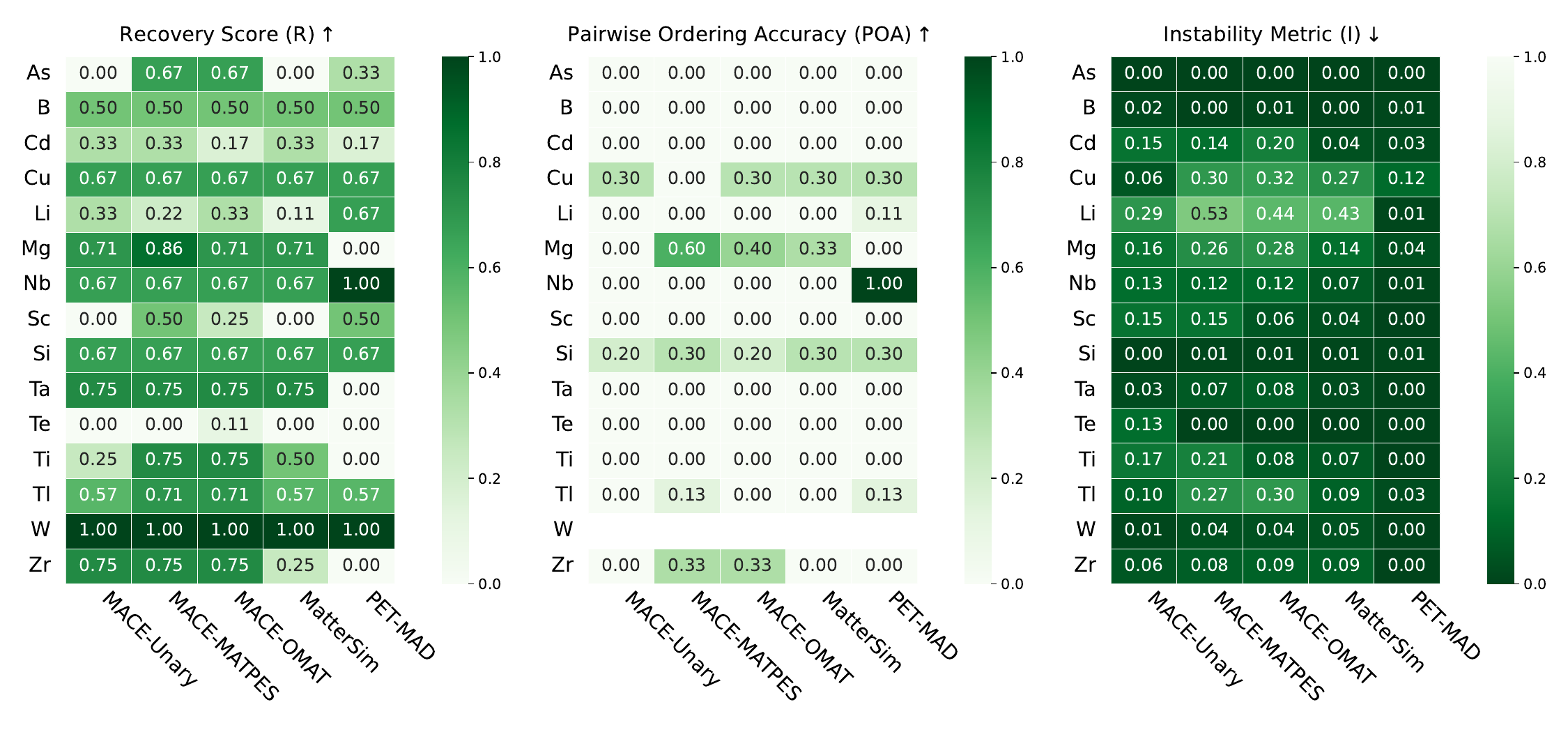}
\caption{Heatmap representation of Minima Hopping benchmarking metrics across elements and models. Panels show recovery score (R), pairwise ordering accuracy (POA), and instability metric (I). Higher R and POA indicate better performance, while lower I corresponds to more stable relaxations.}
\label{fig:heatmap}
\end{figure}

Here, we discuss the results of the extensive structural search performed via MH for the selected elements. The benchmarking study evaluates the ability of each uMLIP to explore the low-energy landscape and accurately reproduce known phases. Specifically, we used MPDB as a reference, focusing on stable and metastable structures with energies less than 100 meV/atom above the convex hull. Although our structural search yielded numerous novel, highly symmetric metastable structures beyond those in MPDB, this study focuses specifically on the models' ability to recover and reproduce known stable and metastable phases. As mentioned before, we quantify this performance using three metrics: Recovery ($R$), pairwise ordering accuracy (POA), and Instability ($I$). A comprehensive comparison of these metrics across all tested uMLIPs is presented in the heatmaps of Fig.~\ref{fig:heatmap}.

As demonstrated by the recovery score metrics, the MACE-MATPES and MACE-OMAT models exhibit comparable performance in exploring the PES of the selected elements. A similar level of consistency is observed between the MACE-Unary and MatterSim models. While PET-MAD generally yields the lowest recovery quality across the dataset, it notably achieves a perfect recovery score of 1.0 for both Nb and W, demonstrating high accuracy for these elements.

Our results for the POA metric indicate that all evaluated models struggle to faithfully replicate the energetic ordering of the identified minima structures. However, it should be noted that this represents an exceptionally tight criterion for uMLIPs. The inherent energy resolution and predictive accuracy of current models often fall short of the precision required to resolve the subtle energetic ordering typical of complex PESs.

Finally, analysis of the instability metrics highlights the superior computational robustness of PET-MAD. By providing a smoother PES, the model achieves higher relaxation efficiency and a lower rate of trajectory failures. This allows it to successfully converge highly distorted configurations that frequently lead to numerical instabilities in alternative model architectures during extensive global minima searches. Comprehensive details of the MH calculations for all models across the benchmarked elements are provided in the Supplementary Material (see Tables S2-S6)~\cite{sup_mat}.

\begin{figure}[H]
\centering
\includegraphics[width=0.49\columnwidth]{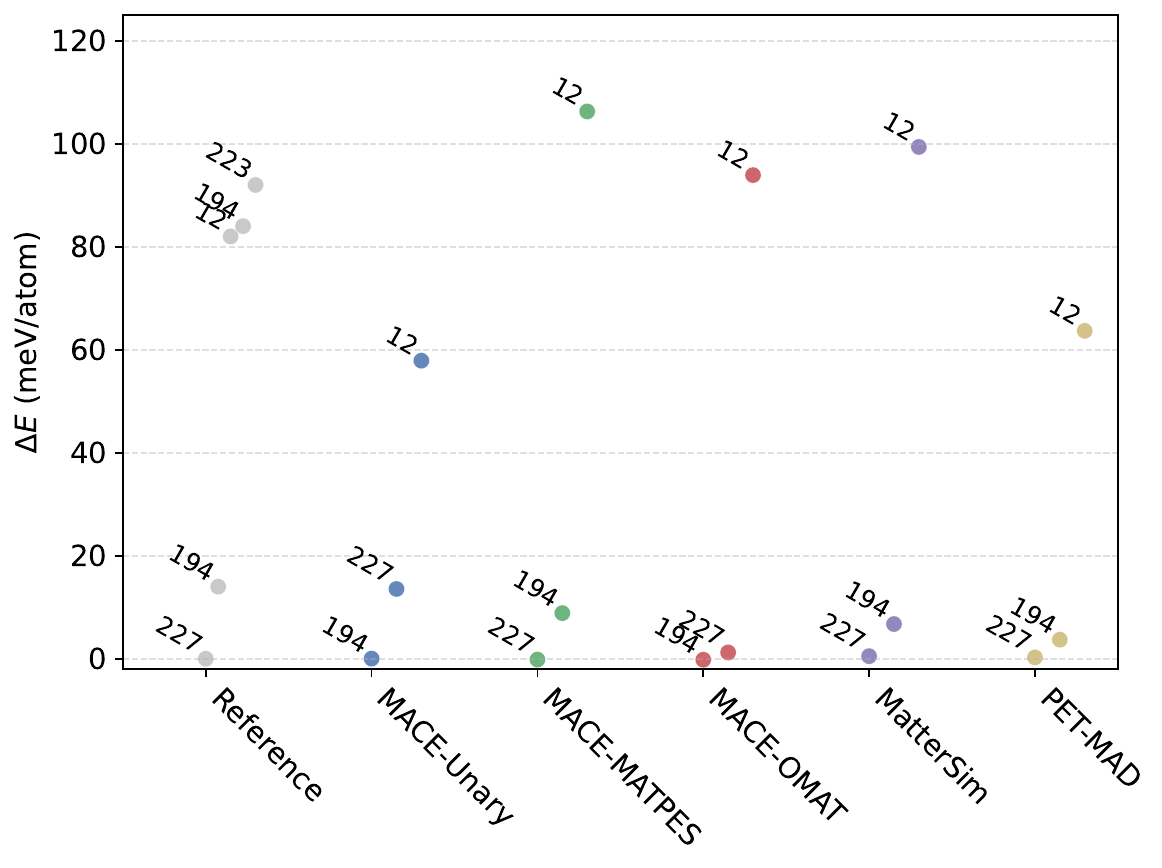}
\includegraphics[width=0.49\linewidth]{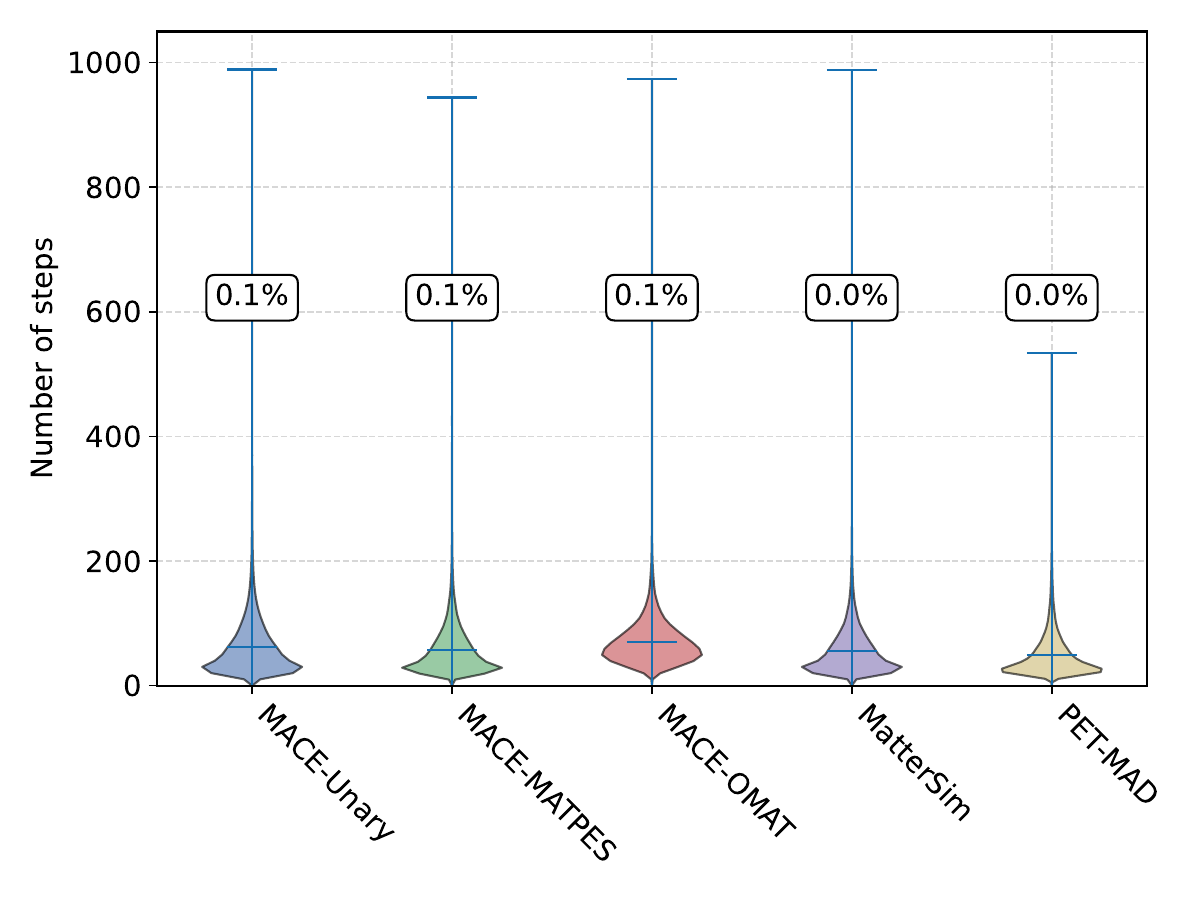}
\caption{Left: Comparison of MH results for Si using uMLIPs with respect to MP reference data. Right: Violin plots showing the distribution of optimization-step counts during local geometry relaxations in the MH trajectories. Box annotations report the failure rate of structural relaxations.}
\label{fig:mh_Si}
\end{figure}

To evaluate the models' performance in global optimization, we present the MH results for Silicon (Si) as a representative case study. Fig.~\ref{fig:mh_Si} illustrates the low-energy structures identified by each model, where the y-axis represents the energy difference ($\Delta E$) relative to each model’s respective GM. These findings are compared against data from MPDB to assess structural recovery and the accuracy of energetic ordering. Notably, all models successfully identify the GM structure (space group Fd$\bar{3}$m1, $227$), as well as the second metastable phase, a hexagonal structure (P6$_3$/mmc) located $14$ meV/atom above the GM at the level of DFT. Overall, the MACE-MATPES, MatterSim, and PET-MAD models reasonably reproduce the energetic ordering of these lowest-energy phases for Si.

The computational efficiency of these global searches is captured by the distribution of optimization-step counts during the local geometry relaxation phases of the MH trajectories (see violin plots in Fig.~\ref{fig:mh_Si}). For Si, PET-MAD exhibits a significantly narrower and lower distribution of optimization steps, indicating a smoother PES that facilitates rapid convergence to local minima. In this case, the other models demonstrate comparable numerical robustness, with nearly a zero-percent failure rate during geometry optimization.

\begin{figure}[H]
\centering
\includegraphics[width=0.49\columnwidth]{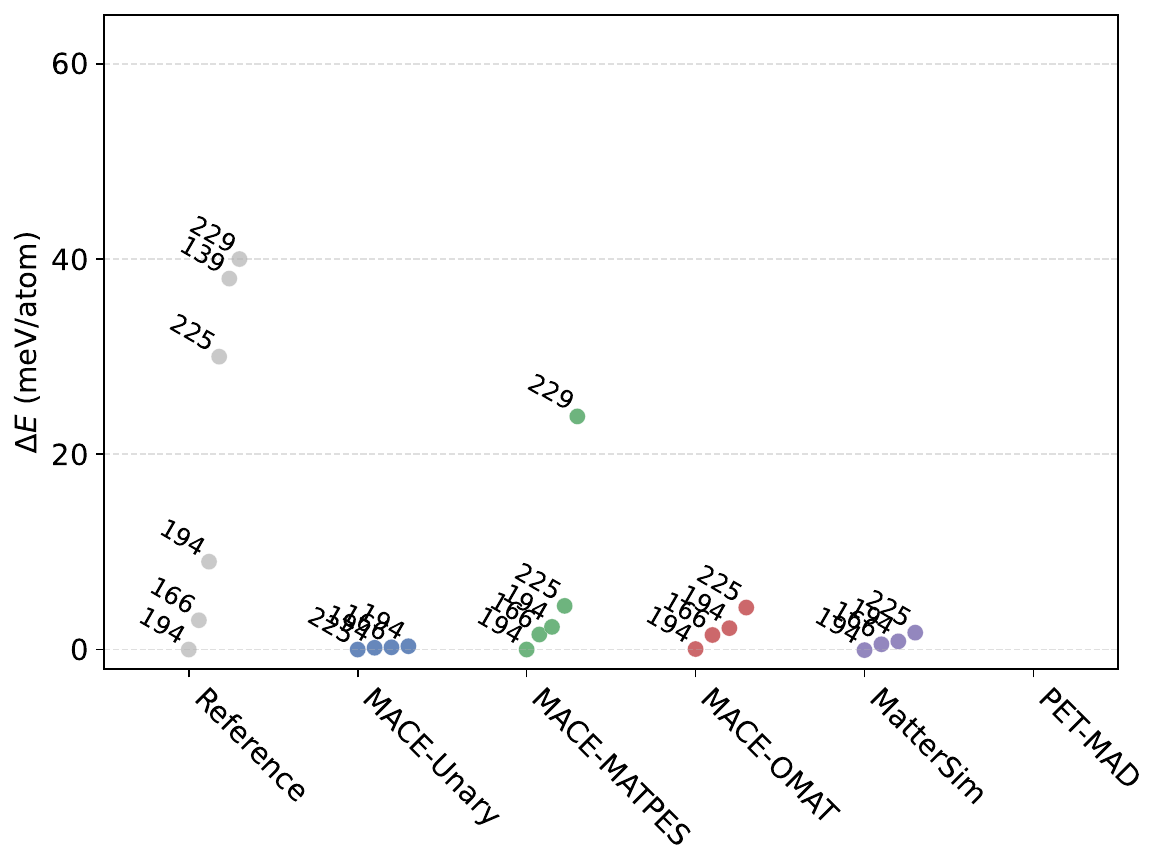}
\includegraphics[width=0.49\linewidth]{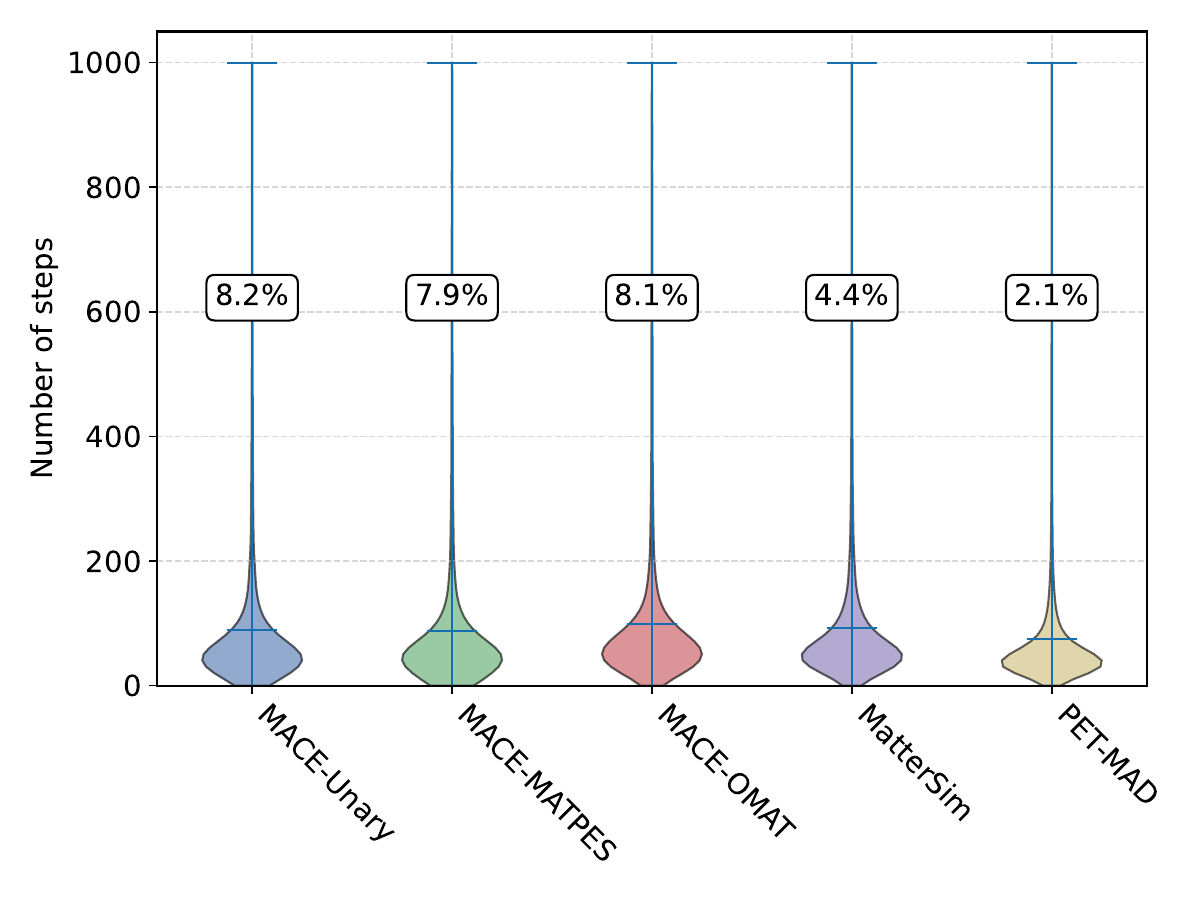}
\caption{Same as Fig.~\ref{fig:mh_Si} for Mg.}
\label{fig:mh_mg}
\end{figure}

Crucially, our results suggest that there is not necessarily a direct correlation between the smoothness of the PES and the quality or accuracy of the discovered minima. While a smoother PES, as seen in the PET-MAD results, facilitates faster local convergence and higher efficiency in terms of computational steps, it does not inherently guarantee a more accurate global energetic landscape. This disconnect is clearly exemplified by the cases of Tellurium (Te) and Magnesium (Mg). For Te, although the violin plots indicate high efficiency and numerical stability across all models, none were able to successfully identify the known minima structures. Conversely, for Mg, the violin plots reveal a high failure rate during local optimizations. Despite this numerical instability, the MACE-based and MatterSim models successfully recovered all known minima, even if the final energetic ordering remained incorrect (see Fig.~\ref{fig:mh_mg}).

\clearpage

These findings highlight a significant insight for model benchmarking: search efficiency (smoothness) and structural fidelity (accuracy) are distinct performance metrics. A model may be numerically well-behaved but lack the physical resolution to find specific phases, while a more ``rugged'' model may be harder to converge but ultimately more successful in navigating complex structural landscapes. Detailed optimization statistics and energy-ordering plots for the full suite of elements are provided in the Supplementary Material (see Figs. S1-S19 and Tables S2-S6).

\section{Summary}
This study presents a robust benchmark of five state-of-the-art uMLIPs across diverse elemental systems. By combining near-equilibrium EOS metrics with global structural exploration via MH, we elucidate both the impressive generalization of current foundation models and their inherent limitations. Our results demonstrate that while these models are increasingly mature, their performance remains sensitive to chemical domain-specific biases and the topological characteristics of the predicted PES.

A primary finding is the significant variation in accuracy across different chemical groups. While MACE-OMAT and MatterSim excel in predicting the properties of transition metals, they underperform on Group 1 alkali metals, a domain where specialized models like MACE-Unary and MACE-MATPES prove more robust. Furthermore, despite successfully identifying many global minima, most models struggle with POA, highlighting a persistent difficulty in resolving the subtle energetic ordering of metastable phases that lie near the models' intrinsic error margins.

Consequently, our findings demonstrate that the quality and accuracy of current uMLIPs are highly promising, particularly given their ability to provide a unified description across diverse elemental systems without the need for system-specific parameterization. However, despite this progress, significant development is still required to ensure that these models provide a consistently accurate PES even for chemically simple systems. While uMLIPs serve as powerful foundational tools for broad exploration, achieving high-fidelity results for more complex structural landscapes remains a challenge. Future advancements must focus on enhancing the intrinsic accuracy of these global models for general applications, while recognizing that task-specific fine-tuning remains essential for resolving the intricate details of complex materials systems.

\section*{Data availability statement}
All benchmark inputs, structure-matching parameters, and analysis scripts will be made publicly available upon publication. The data that support the findings of this study are available from the corresponding author upon reasonable request. The dataset used for training the MACE-Unary model is publicly available and can be accessed via the GitLab repository \url{https://codebase.helmholtz.cloud/casus-datalad/atomistic_foundation_model.datalad/project} and via the interactive metadata inventory
\url{https://data.casus.science/link/elements}.

\section*{Acknowledgements}
The authors gratefully acknowledge the computing time provided to them on the high-performance computer Noctua2 at the NHR Center PC2. This is funded by the Federal Ministry of Education and Research and the state governments participating on the basis of the resolutions of the GWK for the national high-performance computing at universities (www.nhr-verein.de/unsere-partner). The computations for this research were performed using computing resources under project hpc-prf-abtop. The authors would like to acknowledge the European Union’s Just Transition Fund (JTF), administered by the Sächsische Aufbaubank (SAB), under the InfraProNet Research 2021–2027 programme.

\section*{Author contributions statement}
H.M. and T.D.K conceived the idea for this study. H.T. and H.M. conducted the calculations and analyzed the results of the large-scale simulations. A.K. created the data repository and contributed to the writing of the manuscript. All authors contributed in manuscript writing and reviewed the final manuscript.

\section*{Supporting information available}
The Supporting Information contains additional figures.


%

\bibliography{main.bib}

\end{document}



\title[Benchmarking Universal Machine Learning Interatomic Potentials on Elemental Systems]{
Supporting Information for: Benchmarking Universal Interatomic Potentials on Elemental Systems}

%
%

\author[1,2]{\fnm{Hossein} \sur{Tahmasbi}}
\author[1,2]{\fnm{Andreas} \sur{Kn\"upfer}}
\author[1,2]{\fnm{Thomas D.} \sur{K\"uhne}}
\author*[1,2]{\fnm{Hossein} \sur{Mirhosseini}}\email{h.mirhosseini@hzdr.de}
\affil[1]{Center for Advanced Systems Understanding (CASUS), D-02826 G\"orlitz, Germany}

\affil[2]{Helmholtz-Zentrum Dresden-Rossendorf (HZDR), D-01328 Dresden, Germany}

\maketitle

\section{Minima Hopping}
\subsection{MH metrics}

\begin{table}[ht]
\centering
\footnotesize
\begin{tabular*}{\textwidth}{@{\extracolsep\fill}llllllllllllllll}
\toprule%
 & \multicolumn{3}{@{}c@{}}{MACE-Unary}
 & \multicolumn{3}{@{}c@{}}{MACE-MATPES}
 & \multicolumn{3}{@{}c@{}}{MACE-OMAT}
 & \multicolumn{3}{@{}c@{}}{MatterSim}
 & \multicolumn{3}{@{}c@{}}{PET-MAD} \\
\cmidrule{2-15}
elements & $R$ $\uparrow$ & POA $\uparrow$ & I $\downarrow$
& $R$ $\uparrow$ & POA $\uparrow$ & I $\downarrow$
& $R$ $\uparrow$ & POA $\uparrow$ & I $\downarrow$
& $R$ $\uparrow$ & POA $\uparrow$ & I $\downarrow$
& $R$ $\uparrow$ & POA $\uparrow$ & I $\downarrow$ \\
\midrule
As (2) & 0.00 & 0.00 & 0.00 & 0.67 & 0.00 & 0.00 & 0.67 & 0.00 & 0.00 & 0.00 & 0.00 & 0.00 & 0.33 & 0.00 & 0.00 \\
B  (3) & 0.50 & 0.00 & 0.02 & 0.50 & 0.00 & 0.00 & 0.50 & 0.00 & 0.01 & 0.50 & 0.00 & 0.00 & 0.50 & 0.00 & 0.01 \\
Cd (5) & 0.33 & 0.00 & 0.15 & 0.33 & 0.00 & 0.14 & 0.17 & 0.00 & 0.20 & 0.33 & 0.00 & 0.04 & 0.17 & 0.00 & 0.03 \\
Cu (5) & 0.67 & 0.30 & 0.06 & 0.67 & 0.00 & 0.30 & 0.67 & 0.30 & 0.32 & 0.67 & 0.30 & 0.27 & 0.67 & 0.30 & 0.12 \\
Li (8) & 0.33 & 0.00 & 0.29 & 0.22 & 0.00 & 0.53 & 0.33 & 0.00 & 0.44 & 0.11 & 0.00 & 0.43 & 0.67 & 0.107 & 0.01 \\
Mg (6) & 0.71 & 0.00 & 0.16 & 0.86 & 0.60 & 0.26 & 0.71 & 0.40 & 0.28 & 0.71 & 0.333 & 0.14 & 0.00 & 0.00 & 0.04 \\
Nb (2) & 0.67 & 0.00 & 0.13 & 0.67 & 0.00 & 0.12 & 0.67 & 0.00 & 0.12 & 0.67 & 0.00 & 0.07 & 1.00 & 1.00 & 0.01 \\
Sc (3) & 0.00 & 0.00 & 0.15 & 0.50 & 0.00 & 0.15 & 0.25 & 0.00 & 0.06 & 0.00 & 0.00 & 0.04 & 0.50 & 0.00 & 0.00 \\
Si (5) & 0.67 & 0.20 & 0.00 & 0.67 & 0.30 & 0.01 & 0.67 & 0.20 & 0.01 & 0.67 & 0.30 & 0.01 & 0.67 & 0.30 & 0.01 \\
Ta (3) & 0.75 & 0.00 & 0.03 & 0.75 & 0.00 & 0.07 & 0.75 & 0.00 & 0.08 & 0.75 & 0.00 & 0.03 & 0.00 & 0.00 & 0.00 \\
Te (8) & 0.00 & 0.00 & 0.13 & 0.00 & 0.00 & 0.00 & 0.11 & 0.00 & 0.00 & 0.00 & 0.00 & 0.00 & 0.00 & 0.00 & 0.00 \\
Ti (3) & 0.25 & 0.00 & 0.17 & 0.75 & 0.00 & 0.21 & 0.75 & 0.00 & 0.08 & 0.50 & 0.00 & 0.07 & 0.00 & 0.00 & 0.00 \\
Tl (6) & 0.57 & 0.00 & 0.10 & 0.71 & 0.133 & 0.27 & 0.71 & 0.00 & 0.30 & 0.57 & 0.00 & 0.09 & 0.57 & 0.133 & 0.03 \\
W (1)  & 1.00 & $\mathrm{NaN}$ & 0.01 & 1.00 & $\mathrm{NaN}$ & 0.04 & 1.00 & $\mathrm{NaN}$ & 0.04 & 1.00 & $\mathrm{NaN}$ & 0.05 & 1.00 & $\mathrm{NaN}$ & 0.00 \\
Zr (3) &0.75 & 0.00 & 0.06 & 0.75 & 0.333 & 0.08 & 0.75 & 0.333 & 0.09 & 0.25 & 0.00 & 0.09 & 0.00 & 0.00 & 0.00 \\
\botrule
\end{tabular*}
\caption{\label{tab:mh_combined}
Summary of combined recovery ($R$), pairwise ordering accuracy (POA), and instability ($I$) metrics for the MH task across the benchmark elements. The number of MP reference minima is indicated in parentheses next to each element.}
\end{table}

\clearpage

\subsection{MH Performance Summary across all models}

\subsubsection{MACE-Unary-PBE-0}
\begin{table}[ht]
\centering
\begin{tabular*}{\textwidth}{@{\extracolsep\fill}lllllll}
\toprule%
         & \#optimization & \#failed & \#visited & \#accepted & \#catas. & I (\%) \\
elements & attempts & optimization  & minima & minima & latt. exp. & \\
\midrule
As & 43970 & 1 & 32798 & 15301 & 0 & 0.0 \\
B & 26156 & 0 & 16259 & 7280 & 357 & 2.2 \\
Cd & 35764 & 3060 & 28811 & 12975 & 1846 & 14.96 \\
Cu & 27749 & 1217 & 18657 & 6541 & 218 & 5.55 \\
Li & 25851 & 4337 & 13047 & 4608 & 1644 & 29.38 \\
Mg & 23633 & 1909 & 17134 & 7170 & 1324 & 15.81 \\
Nb & 19277 & 329 & 9075 & 1775 & 1023 & 12.98 \\
Sc & 16151 & 358 & 9927 & 6296 & 1297 & 15.28 \\
Si & 39932 & 21 & 28665 & 12894 & 6 & 0.07 \\
Ta & 28963 & 525 & 16491 & 5202 & 165 & 2.81 \\
Te & 43691 & 8 & 32292 & 15442 & 4295 & 13.32 \\
Ti & 23445 & 2299 & 16758 & 7067 & 1276 & 17.42 \\
Tl & 31844 & 1189 & 23431 & 10461 & 1574 & 10.45 \\
W & 22223 & 35 & 12968 & 4094 & 44 & 0.5 \\
Zr & 28398 & 504 & 22027 & 8320 & 841 & 5.59 \\
 \botrule
\end{tabular*}
\caption{\label{tab:mh_unary}
Details of MH calculations with MACE-Unary-PBE-0 across the benchmark elements.}
\end{table}
\clearpage
\subsubsection{MACE-MATPES-PBE-0}
\begin{table}[ht]
\centering
\begin{tabular*}{\textwidth}{@{\extracolsep\fill}lllllll}
\toprule%
         & \#optimization & \#failed & \#visited & \#accepted & \#catas. & I (\%) \\
elements & attempts & optimization  & minima & minima & latt. exp. & \\
\midrule
As & 45217 & 1 & 32699 & 14951 & 0 & 0.0 \\
B & 25880 & 0 & 16260 & 6983 & 42 & 0.26 \\
Cd & 39524 & 473 & 33190 & 14302 & 4393 & 14.43 \\
Cu & 32747 & 4361 & 23669 & 9093 & 3903 & 29.81 \\
Li & 36026 & 9961 & 21968 & 8421 & 5540 & 52.87 \\
Mg & 30213 & 2466 & 24190 & 9306 & 4333 & 26.07 \\
Nb & 45495 & 1513 & 28177 & 9029 & 2517 & 12.26 \\
Sc & 36861 & 2077 & 29115 & 11679 & 2810 & 15.29 \\
Si & 45709 & 48 & 28586 & 10878 & 209 & 0.84 \\
Ta & 32711 & 765 & 19414 & 6168 & 885 & 6.9 \\
Te & 47462 & 20 & 32766 & 14420 & 7 & 0.06 \\
Ti & 34730 & 2909 & 24539 & 9445 & 3051 & 20.81 \\
Tl & 39997 & 6851 & 32048 & 13805 & 3297 & 27.42 \\
W & 24261 & 195 & 13649 & 4165 & 419 & 3.87 \\
Zr & 38698 & 1362 & 30818 & 12950 & 1324 & 7.82 \\
 \botrule
\end{tabular*}
\caption{\label{tab:mh_matpes}
Details of MH calculations with MACE-MATPES-PBE-0 across the benchmark elements.}
\end{table}
\clearpage
\subsubsection{MACE-OMAT-0}
\begin{table}[ht]
\centering
\begin{tabular*}{\textwidth}{@{\extracolsep\fill}lllllll}
\toprule%
         & \#optimization & \#failed & \#visited & \#accepted & \#catas. & I (\%) \\
elements & attempts & optimization  & minima & minima & latt. exp. & \\
\midrule
As & 46253 & 6 & 32566 & 15198 & 2 & 0.02 \\
B & 20352 & 0 & 12184 & 5052 & 181 & 1.49 \\
Cd & 39717 & 251 & 33317 & 13970 & 6360 & 19.72 \\
Cu & 34400 & 3693 & 24782 & 8603 & 5189 & 31.67 \\
Li & 35171 & 7627 & 21968 & 8470 & 4982 & 44.36 \\
Mg & 30460 & 2403 & 23984 & 8944 & 4941 & 28.49 \\
Nb & 45593 & 1229 & 28229 & 9060 & 2539 & 11.69 \\
Sc & 38465 & 654 & 29802 & 12132 & 1203 & 5.74 \\
Si & 46079 & 24 & 28530 & 10760 & 202 & 0.76 \\
Ta & 31626 & 802 & 19431 & 6458 & 1131 & 8.36 \\
Te & 48505 & 0 & 32762 & 14772 & 3 & 0.01 \\
Ti & 31724 & 655 & 25820 & 11508 & 1476 & 7.78 \\
Tl & 39228 & 6462 & 31166 & 13753 & 4090 & 29.6 \\
W & 23958 & 132 & 13736 & 4216 & 436 & 3.73 \\
Zr & 37713 & 1236 & 30683 & 13322 & 1790 & 9.11 \\
 \botrule
\end{tabular*}
\caption{\label{tab:mh_omat}
Details of MH calculations with MACE-OMAT-0 across the benchmark elements.}
\end{table}
\clearpage
\subsubsection{MatterSim}
\begin{table}[ht]
\centering
\begin{tabular*}{\textwidth}{@{\extracolsep\fill}lllllll}
\toprule%
         & \#optimization & \#failed & \#visited & \#accepted & \#catas. & I (\%) \\
elements & attempts & optimization  & minima & minima & latt. exp. & \\
\midrule
As & 41706 & 0 & 32541 & 16352 & 40 & 0.12 \\
B & 25581 & 6 & 16361 & 6875 & 0 & 0.02 \\
Cd & 37938 & 383 & 32406 & 15550 & 1124 & 4.48 \\
Cu & 34830 & 2488 & 25023 & 9235 & 5044 & 27.3 \\
Li & 37612 & 4487 & 23510 & 8516 & 7345 & 43.17 \\
Mg & 32401 & 1418 & 25471 & 10419 & 2493 & 14.16 \\
Nb & 44178 & 671 & 27673 & 9348 & 1403 & 6.59 \\
Sc & 34165 & 532 & 29248 & 13902 & 719 & 4.02 \\
Si & 45423 & 16 & 28592 & 10972 & 251 & 0.91 \\
Ta & 31071 & 203 & 18998 & 6600 & 425 & 2.89 \\
Te & 41895 & 3 & 32727 & 16424 & 17 & 0.06 \\
Ti & 30071 & 881 & 25759 & 13039 & 1104 & 7.22 \\
Tl & 38395 & 801 & 32564 & 14552 & 2298 & 9.14 \\
W & 23730 & 30 & 13800 & 4375 & 670 & 4.98 \\
Zr & 37073 & 1538 & 31086 & 14204 & 1524 & 9.05 \\
 \botrule
\end{tabular*}
\caption{\label{tab:mh_mattersim}
Details of MH calculations with MatterSim across the benchmark elements.}
\end{table}
\clearpage
\subsubsection{PET-MAD}
\begin{table}[ht]
\centering
\begin{tabular*}{\textwidth}{@{\extracolsep\fill}lllllll}
\toprule%
         & \#optimization & \#failed      & \#visited       & \#accepted & \#catas. & I (\%) \\
elements & attempts       & optimization  & minima & minima & latt. exp. & \\
\midrule
As & 36079 & 0 & 32628 & 17853 & 8 & 0.02 \\
B & 23309 & 15 & 16099 & 7515 & 177 & 1.16 \\
Cd & 36290 & 36 & 32559 & 16696 & 1001 & 3.17 \\
Cu & 31144 & 2415 & 25894 & 12125 & 1068 & 11.88 \\
Li & 31227 & 205 & 28420 & 14362 & 189 & 1.32 \\
Mg & 28585 & 598 & 26159 & 13360 & 477 & 3.92 \\
Nb & 31625 & 130 & 28759 & 15812 & 92 & 0.73 \\
Sc & 33310 & 20 & 29955 & 15412 & 5 & 0.08 \\
Si & 36797 & 0 & 28526 & 14381 & 182 & 0.64 \\
Ta & 21588 & 17 & 19704 & 10619 & 16 & 0.16 \\
Te & 34831 & 11 & 32565 & 18104 & 1 & 0.03 \\
Ti & 29357 & 74 & 26258 & 13734 & 46 & 0.43 \\
Tl & 36134 & 1 & 32549 & 16668 & 954 & 2.93 \\
W & 16445 & 47 & 13838 & 7725 & 27 & 0.48 \\
Zr & 33129 & 26 & 31081 & 16573 & 30 & 0.18 \\
\botrule
\end{tabular*}
\caption{\label{tab:mh_petmad}
Details of MH calculations with PET-MAD across the benchmark elements.}
\end{table}
\clearpage
\subsection{MH results for all elements}

\begin{figure}[H]
\centering
\includegraphics[width=0.49\columnwidth]{figs_SI/As_MH.pdf}
\includegraphics[width=0.49\linewidth]{figs_SI/violin_plot_OPT_As.pdf}
\caption{Left: Comparison of MH results for As using uMLIPs with respect to MP reference data. Right: Violin plots showing the distribution of optimization-step counts during local geometry relaxations in the MH trajectories. Box annotations report the failure rate of structural relaxations.}
\label{fig:mh_As}
\end{figure}


\begin{figure}[H]
\centering
\includegraphics[width=0.49\columnwidth]{figs_SI/B_MH.pdf}
\includegraphics[width=0.49\linewidth]{figs_SI/violin_plot_OPT_B.pdf}
\caption{Same as Fig.~\ref{fig:mh_As} for B.}
\label{fig:mh_B}
\end{figure}

\begin{figure}[H]
\centering
\includegraphics[width=0.49\columnwidth]{figs_SI/Cd_MH.pdf}
\includegraphics[width=0.49\linewidth]{figs_SI/violin_plot_OPT_Cd.pdf}
\caption{Same as Fig.~\ref{fig:mh_As} for Cd.}
\label{fig:mh_Cd}
\end{figure}


\begin{figure}[H]
\centering
\includegraphics[width=0.49\columnwidth]{figs_SI/Cu_MH.pdf}
\includegraphics[width=0.49\linewidth]{figs_SI/violin_plot_OPT_Cu.pdf}
\caption{Same as Fig.~\ref{fig:mh_As} for Cu.}
\label{fig:mh_cu}
\end{figure}
 

\begin{figure}[H]
\centering
\includegraphics[width=0.49\columnwidth]{figs_SI/Li_MH.pdf}
\includegraphics[width=0.49\linewidth]{figs_SI/violin_plot_OPT_Li.pdf}
\caption{Same as Fig.~\ref{fig:mh_As} for Li.}
\label{fig:mh_li}
\end{figure}


\begin{figure}[H]
\centering
\includegraphics[width=0.49\columnwidth]{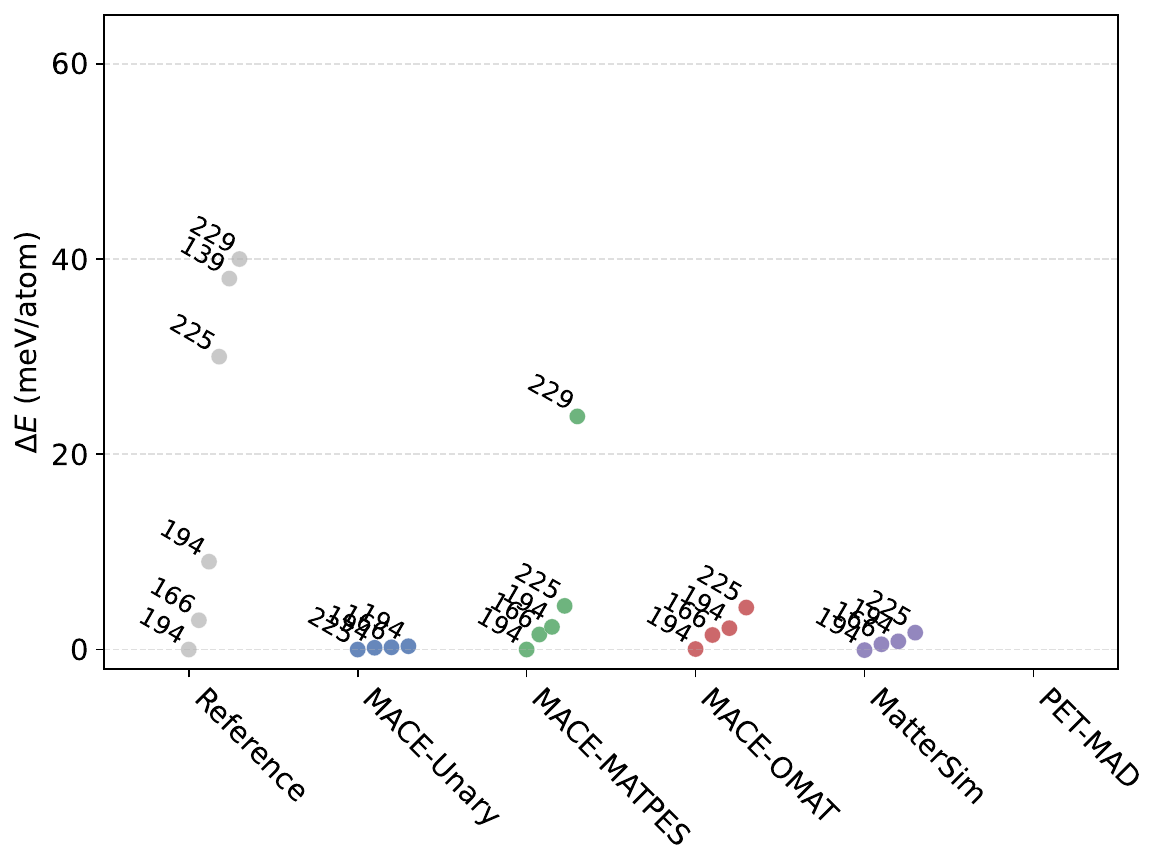}
\includegraphics[width=0.49\linewidth]{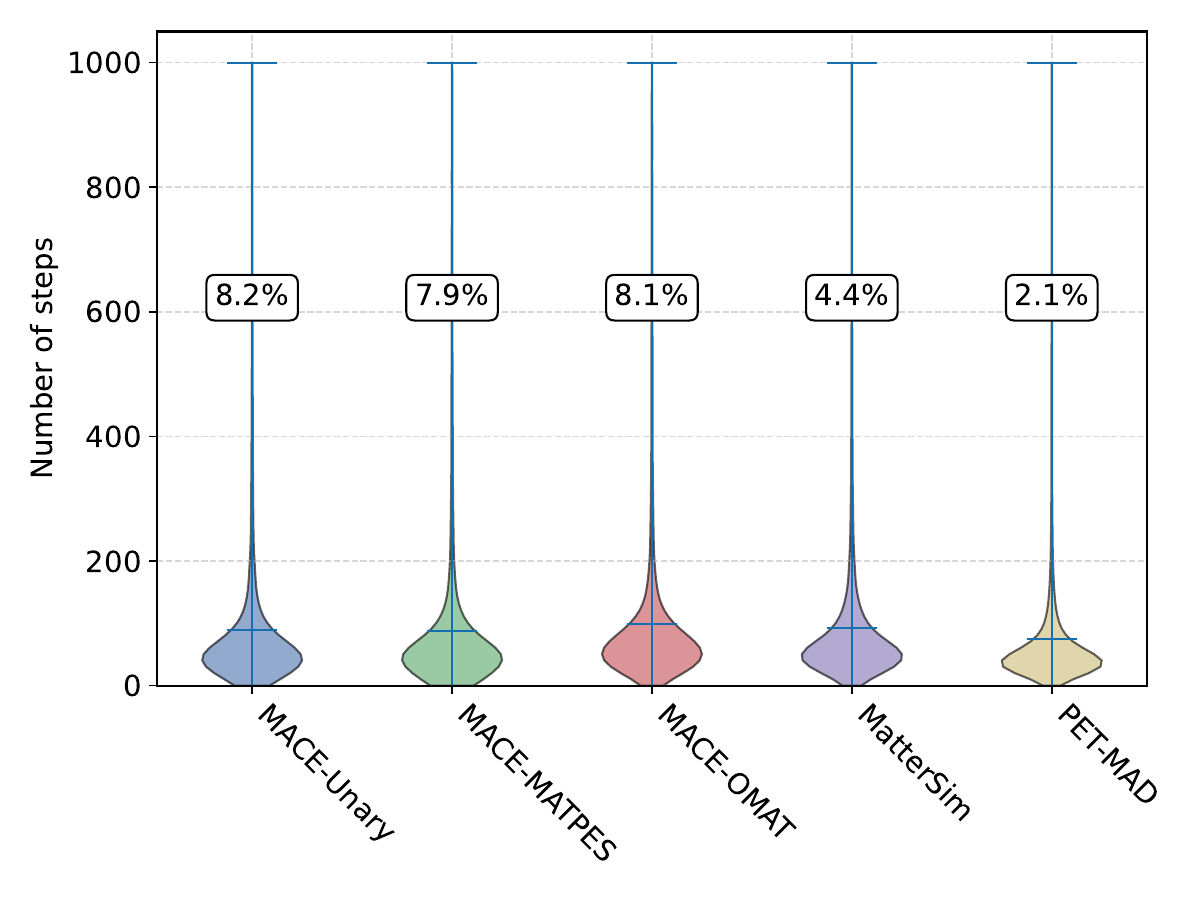}
\caption{Same as Fig.~\ref{fig:mh_As} for Mg.}
\label{fig:mh_mg}
\end{figure}

\begin{figure}[H]
\centering
\includegraphics[width=0.49\columnwidth]{figs_SI/Nb_MH.pdf}
\includegraphics[width=0.49\linewidth]{figs_SI/violin_plot_OPT_Nb.pdf}
\caption{Same as Fig.~\ref{fig:mh_As} for Nb.}
\label{fig:mh_Nb}
\end{figure}

\begin{figure}[H]
\centering
\includegraphics[width=0.49\columnwidth]{figs_SI/Sc_MH.pdf}
\includegraphics[width=0.49\linewidth]{figs_SI/violin_plot_OPT_Sc.pdf}
\caption{Same as Fig.~\ref{fig:mh_As} for Sc.}
\label{fig:mh_Sc}
\end{figure}

\begin{figure}[H]
\centering
\includegraphics[width=0.49\columnwidth]{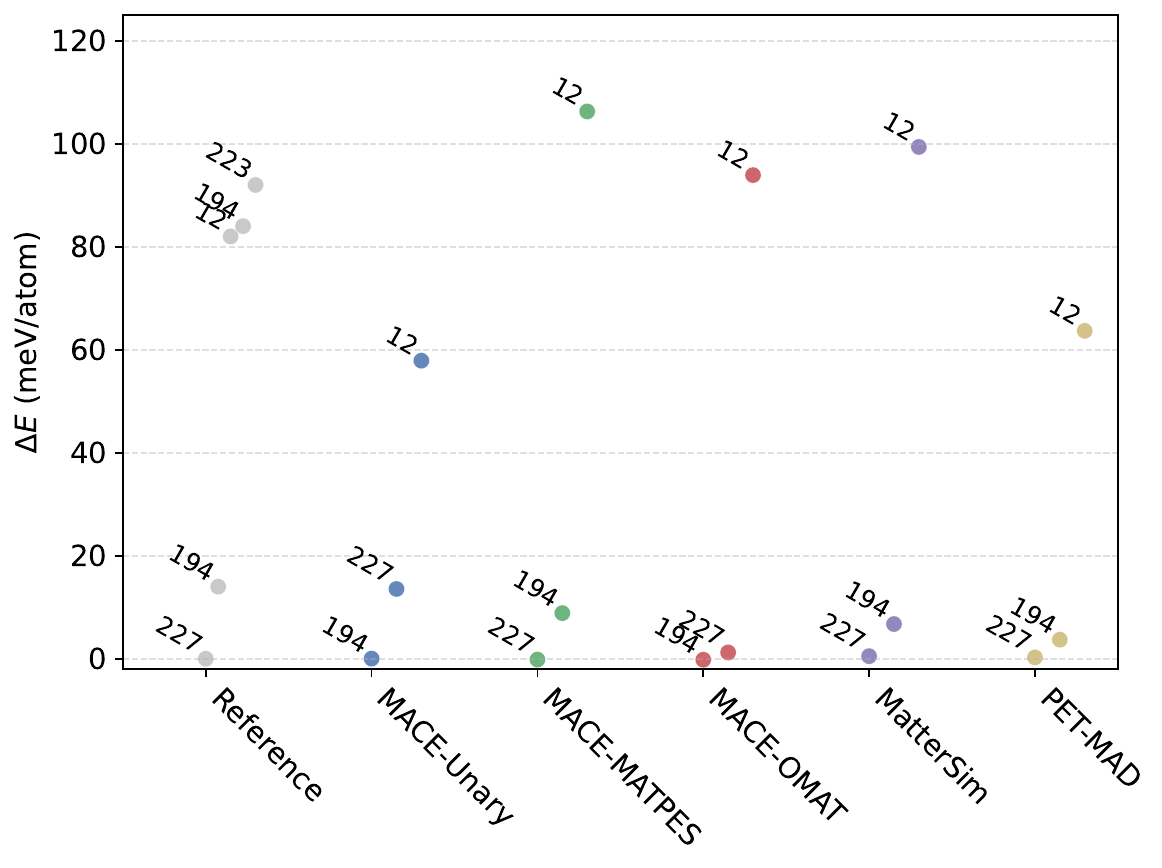}
\includegraphics[width=0.49\linewidth]{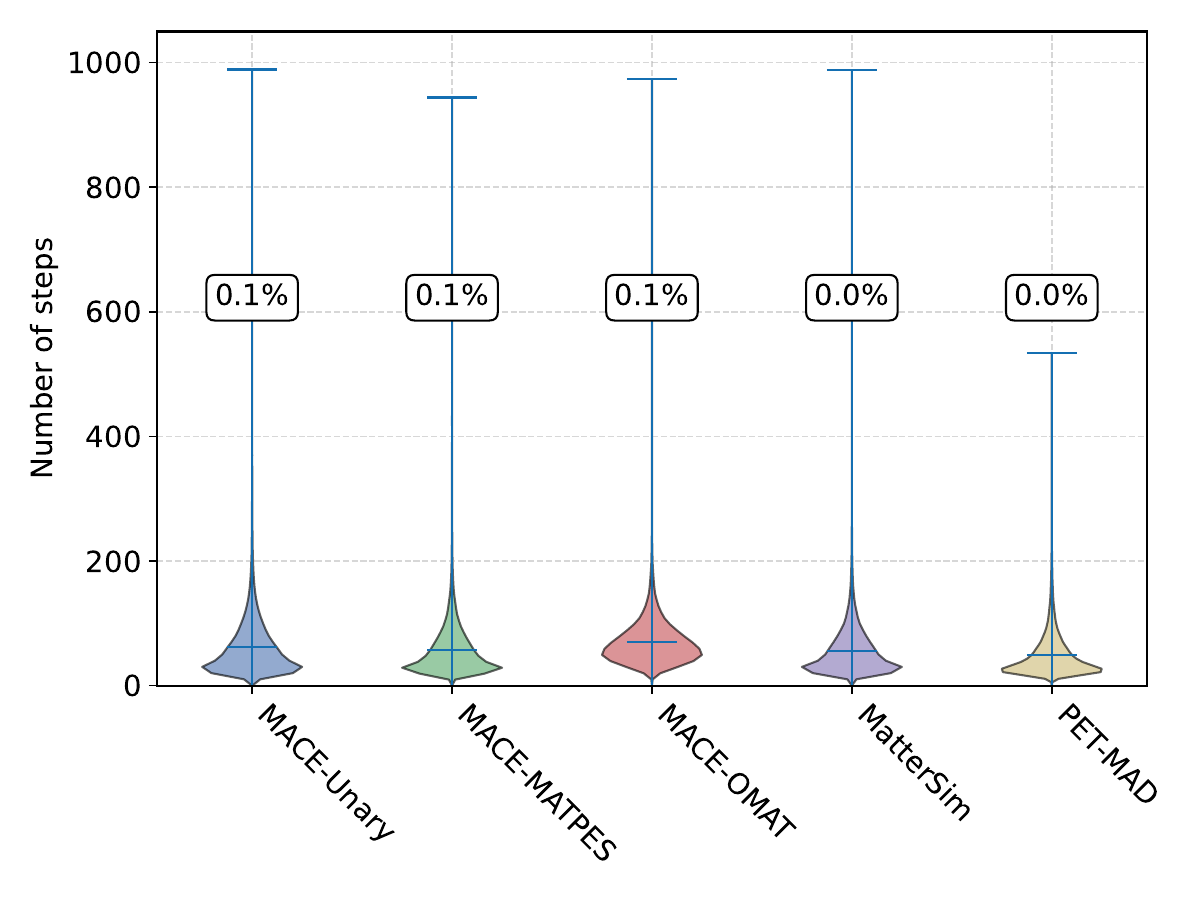}
\caption{Same as Fig.~\ref{fig:mh_As} for Si.}
\label{fig:mh_Si}
\end{figure}


\begin{figure}[H]
\centering
\includegraphics[width=0.49\columnwidth]{figs_SI/Ta_MH.pdf}
\includegraphics[width=0.49\linewidth]{figs_SI/violin_plot_OPT_Ta.pdf}
\caption{Same as Fig.~\ref{fig:mh_As} for Ta.}
\label{fig:mh_Ta}
\end{figure}

\begin{figure}[H]
\centering
\includegraphics[width=0.49\columnwidth]{figs_SI/Te_MH.pdf}
\includegraphics[width=0.49\linewidth]{figs_SI/violin_plot_OPT_Te.pdf}
\caption{Same as Fig.~\ref{fig:mh_As} for Te.}
\label{fig:mh_Te}
\end{figure}


\begin{figure}[H]
\centering
\includegraphics[width=0.49\columnwidth]{figs_SI/Ti_MH.pdf}
\includegraphics[width=0.49\linewidth]{figs_SI/violin_plot_OPT_Ti.pdf}
\caption{Same as Fig.~\ref{fig:mh_As} for Ti.}
\label{fig:mh_ti}
\end{figure}

\begin{figure}[H]
\centering
\includegraphics[width=0.49\columnwidth]{figs_SI/Tl_MH.pdf}
\includegraphics[width=0.49\linewidth]{figs_SI/violin_plot_OPT_Tl.pdf}
\caption{Same as Fig.~\ref{fig:mh_As} for Tl.}
\label{fig:mh_Tl}
\end{figure}

\begin{figure}[H]
\centering
\includegraphics[width=0.49\columnwidth]{figs_SI/W_MH.pdf}
\includegraphics[width=0.49\linewidth]{figs_SI/violin_plot_OPT_W.pdf}
\caption{Same as Fig.~\ref{fig:mh_As} for W.}
\label{fig:mh_W}
\end{figure}

\begin{figure}[H]
\centering
\includegraphics[width=0.49\columnwidth]{figs_SI/Zr_MH.pdf}
\includegraphics[width=0.49\linewidth]{figs_SI/violin_plot_OPT_Zr.pdf}
\caption{Same as Fig.~\ref{fig:mh_As} for Zr.}
\label{fig:mh_Zr}
\end{figure}